\documentclass[10pt,twocolumn]{article} 
\usepackage{simpleConference}
\usepackage{times}
\usepackage{graphicx}
\usepackage{amssymb}
\usepackage{url,hyperref}

\begin{document}

\title{Synthesis and structural/microstructural characteristics of antimony doped tin oxide $(Sn_{1-x}Sb_{x}O_{2-\delta})$}

\author{ \textbf{Rita Singh$^{1}$,  Sushant Gupta$^{2}$ and B. Das$^{3}$} \\
{\normalsize $^{1,2,3}$Department of Physics, University of Lucknow, Lucknow-226 007,
    India} \\ \normalsize{$^{1}$E-mail: ritasingh.singh88@gmail.com}\\
\normalsize{$^{2}$ E-mail: sushant1586@gmail.com}\\
\normalsize{$^{3}$E-mail: bdas226010@gmail.com}
 }

\maketitle
\thispagestyle{empty}

\begin{abstract}
Bulk samples of $(Sn_{1-x}Sb_{x}O_{2-\delta})$ with x = 0.00, 0.10,
0.20, 0.30 are synthesized by solid-state reaction route. Samples
were characterized by X-ray powder diffraction (XRD), scanning
electron microscopy (SEM), transmission electron microscopy (TEM)
and UV-Vis spectroscopy. The x-ray diffraction patterns indicate
that the gross structure/phase of $(Sn_{1-x}$ $Sb_{x}O_{2-\delta})$
do not change with the substitution of antimony (Sb) up to x = 0.30.
The surface morphological examination with SEM revealed the fact
that the grain size in the antimony doped sample is larger than that
of undoped one and hence pores/voids between the grains increase
with Sb concentration up to 0.30. TEM image of undoped sample
indicates that the $SnO_{2}$ grains have diameters ranging from 25
to 120 nm and most grains are in cubic or spherical shape. As
antimony content increases, the nanocubes/spheres are converted into
microcubes/spheres. The reflectance of $Sn_{1-x}Sb_{x}O_{2-\delta}$
samples increases whereas absorbance of these samples decreases with
the increased concentration of antimony (Sb) for the wavelength
range 360 - 800 nm. The energy bandgap of Sb doped - $SnO_{2}$
samples were obtained from optical absorption spectra by UV-Vis
absorption spectroscopy. Upon increasing the Sb concentration the
bandgap of the samples was found to increase from 3.367 eV to 3.558
eV.

\end{abstract}

\section{Introduction}

Metal oxides play a very important role in many areas of chemistry,
physics, and materials science [1-6]. The unique characteristics of
metal oxides make them a very diverse class of materials, with
properties covering almost all aspects of materials science and
solid-state physics. Oxidic materials exhibit fascinating electronic
and magnetic properties, including metallic, semiconducting,
superconducting, or insulating and ferro-, ferri-, or
antiferromagnetic behaviors. In technological applications, oxides
are used in the fabrication of microelectronic circuits \cite{ref7},
capacitors \cite{ref8}, sensors \cite{ref9}, piezoelectric devices
\cite{ref10}, fuel cells \cite{ref11}, semiconductors [12,13],
oxygen generators \cite{ref14}, organic synthetics [15-19], the
manufacture of engineered ceramics \cite{ref20}, coatings for the
passivation of surfaces against corrosion \cite{ref21} and as
catalysts as both the support and active component [22-24]. However,
nanoscale metal oxides are particularly attractive to both pure and
applied researchers because of the great variety of structure and
properties, especially those related to intrinsic size-dependent
properties [11,24-27].\\\\Tin (IV) dioxide (II)
$(SnO_{2})$ is one of the important members of the metal oxide
family and has been extensively studied for wide range of
applications such as gas sensors, transparent conductive oxides,
catalysis, and far-infrared dichromic mirrors. Chemical and thermal
stability, natural off stoichiometery, optical transparency and
possibility of conductivity variation over a wide range makes
$SnO_{2}$ suitable for above mentioned applications
\cite{ref28}.\\\\$SnO_{2}$ has only one stable phase, which
is known as cassiterite (mineral form) or rutile (crystal structure)
with space group of $D_{4}^{th} [P4_{2}/mnm]$. Its unit cell
contains six atoms (two tin and four oxygen), in which tin atoms
occupy the center of a surrounding core composed of six oxygen atoms
placed approximately at the corner of a quasi-regular octahedron
(figure 1).
%%%%%%%%%%%%%%%%%%% Figure 1 %%%%%%%%%%%%%%%%%%%%%%%%%%%%%%%%%%%
\begin{figure}[h]
  \begin{center}
    \includegraphics[width= 8.5cm,height=6.5cm,angle=0]{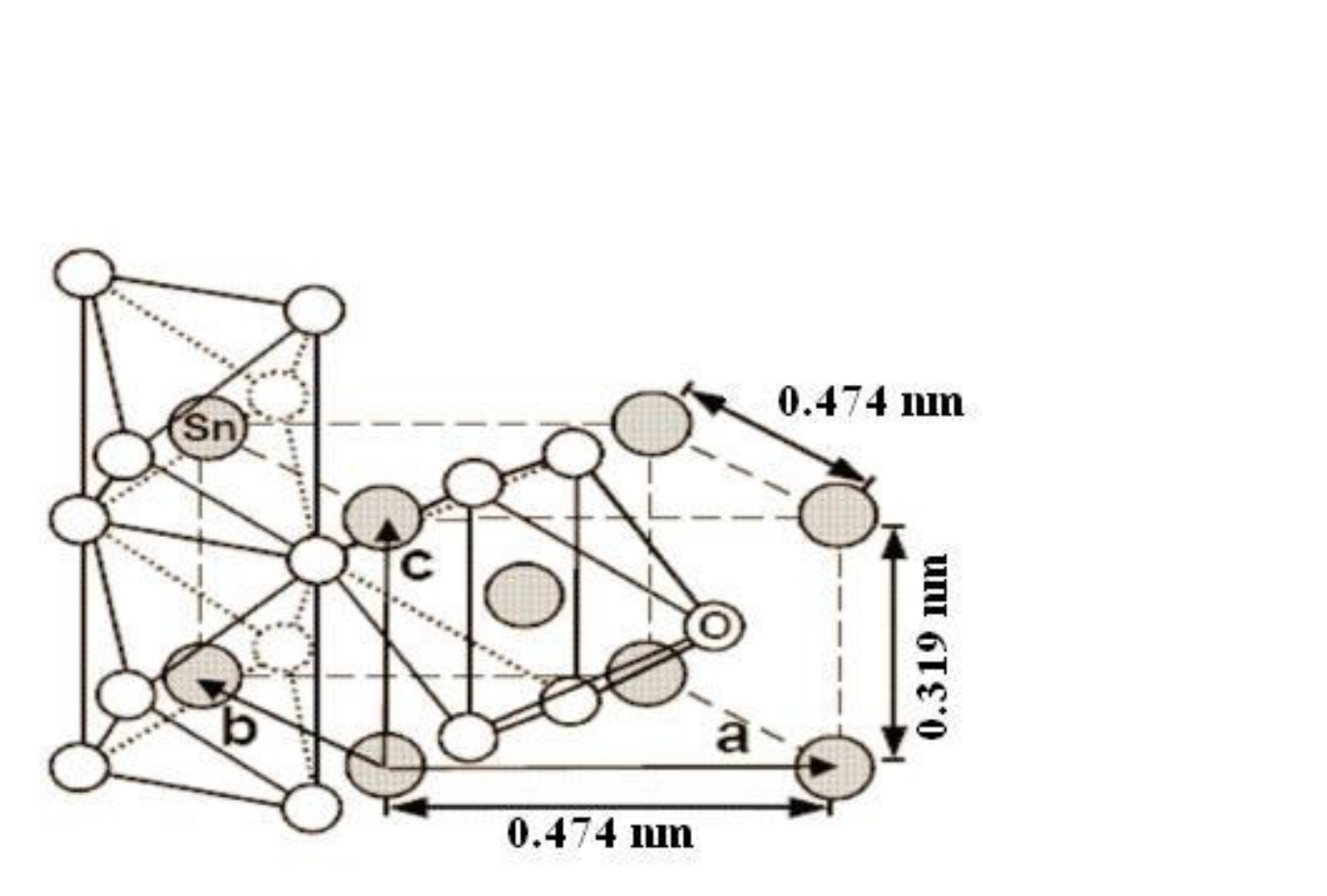}
  \end{center}

  \caption{\small Unit cell of $SnO_{2}$ with four $O^{2-}$ anions and two $Sn^{4+}$ cations. The crystalline structure of $SnO_{2}$
is rutile: Each tin atom is at the centre of six oxygen atoms placed approximately at the corners of a
regular slightly deformed octahedron and three tin atoms approximately at the corners of an equilateral
triangle surround every oxygen atom.}
  \label{fig-label}
\end{figure}
%%%%%%%%%%%%%%%%%%%%%%%%%%%%%%%%%%%%%%%%%%%%%%%%%%%%%%%%%%%%%%%%%
%%%%%%%%%%%%%%%%%%% Figure 2 %%%%%%%%%%%%%%%%%%%%%%%%%%%%%%%%%%%
\begin{figure}[t]
  \begin{center}
    \includegraphics[width= 8.5cm,height=10cm,angle=0]{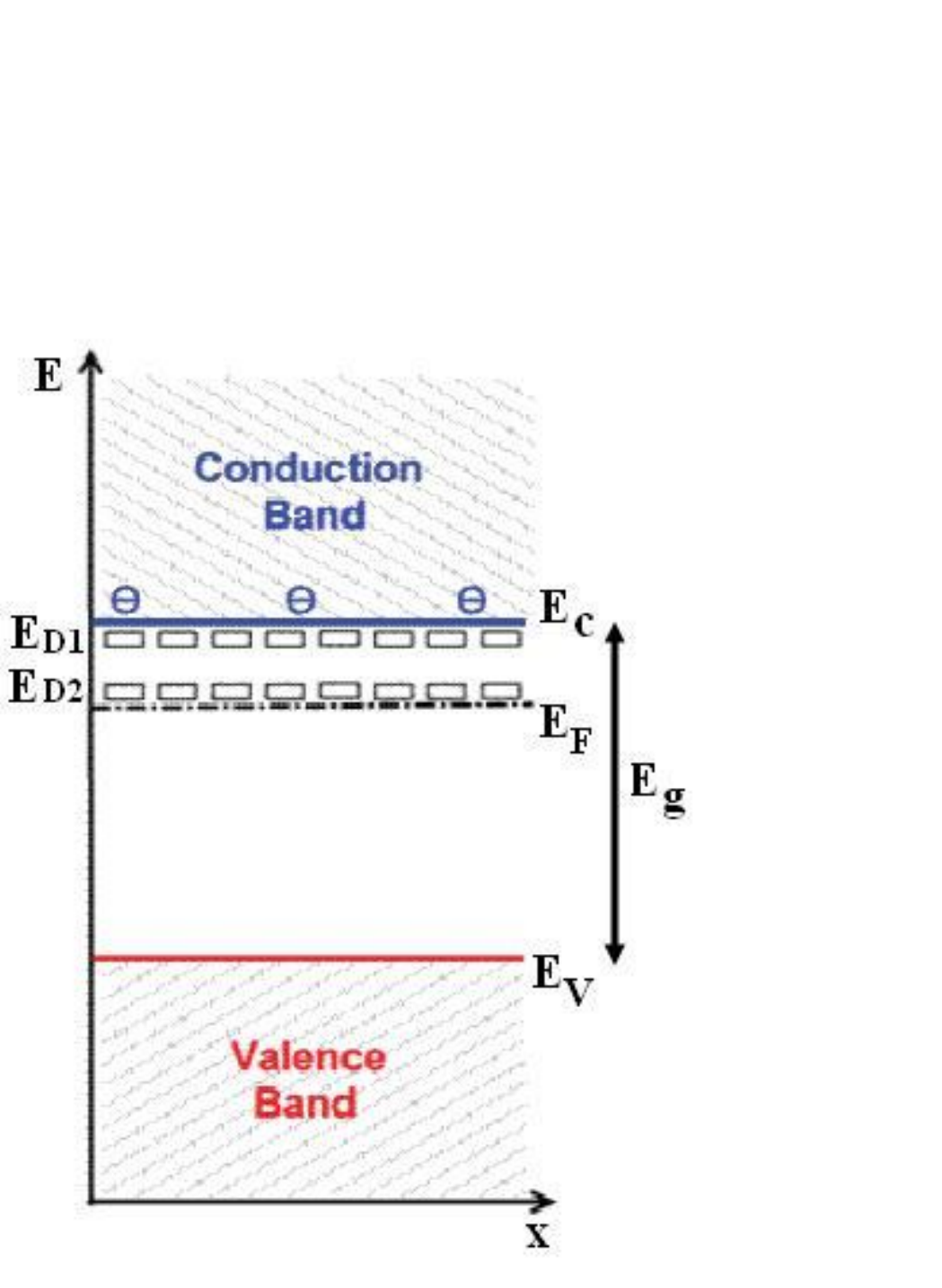}
  \end{center}

  \caption{\small Schematic band diagram of the $SnO_{2}$ bulk. Two
vacancy donor levels $E_{D1}$ and $E_{D2}$ are located 0.03 and 0.15
eV below the conduction band $(E_{C} = 0 eV)$. The bandgap $(E_{g})$
is 3.6 eV.}
 \label{fig-label}
\end{figure}
%%%%%%%%%%%%%%%%%%%%%%%%%%%%%%%%%%%%%%%%%%%%%%%%%%%%%%%%%%%%%%%%%
In case of oxygen atoms, three tin atoms
surround them, approximately forming an equilateral triangle. The
metal atoms ($Sn^{4+}$ cations) are located at (0,0,0) and
($\frac{1}{2},\frac{1}{2},\frac{1}{2}$) positions in the unit cell
and oxygen atoms ($O^{2-}$ anions) at $\pm(u,u,0)$ and
$\pm(\frac{1}{2}u,\frac{1}{2}u,\frac{1}{2})$, where the internal
parameter u takes the value 0.307.\\\\Lattice parameters
are \textbf{a} = \textbf{b} = 0.474 nm and \textbf{c} = 0.319 nm. It
is a wide bandgap degenerate semiconductor with a direct band gap of
approximately 3.6 eV falling in the ultra-violet region, which is an
important feature for many technological applications. The valence
electronic configurations of Sn and O atoms are $5s^{2} 5p^{2}$ and
$2s^{2} 2p^{4}$, respectively. Therefore, while forming the
$SnO_{2}$ molecule, the 5s and 5p electrons of the Sn atom are
partially transferred to two O atoms. Each O atom can accept two
electrons in its 2p orbital to form a stable octet. The energy band
structure of $SnO_{2}$ thus consists of a 5s conduction band and a
2p valence band separated by a forbidden gap.\\\\In its
stoichiometric form $SnO_{2}$ acts as an insulator, but in its
oxygen-deficient form tin dioxide behaves as an n-type
semiconductor. Although the conductivity is thought to be due to
intrinsic defect formation, the exact mechanism is not well
understood. The bulk electronic structure has been investigated by a
number of authors [29-33], although the detailed defect electronic
structure has not yet been studied. Oxygen deficiency may be caused
either by oxygen vacancies or tin interstitials, with reduction of
some Sn(IV) ions to Sn(II) as a possible charge compensation
mechanism as suggested by the observation of a SnO phase in
high-resolution transmission electron microscopy (HRTEM)
\cite{ref34} and fine structure emissions ascribed to Sn(II) in
photoluminescence spectra \cite{ref35}. Electronic conductivity
could then occur due to the mobility of electrons from Sn(II) to
Sn(IV) sites [36, 37].\\\\Experimental data suggests that
the cause of the nonstoichiometry in $SnO_{2}$ is oxygen vacancies
rather than tin interstitials. The measurement of conductivity as a
function of $O_{2}$ partial pressure has shown results consistent
with the oxygen vacancy model [38, 39]. Electron paramagnetic
resonance (EPR) studies have identified a resonance at g = 1.89
associated with singly ionized oxygen vacancies ($V_{O}^{.}$ in
Kr\"{o}ger-Vink notation) [40-42]. This resonance occurs after CO/Ar
treatment of $SnO_{2}$, which results in the formation of
$V_{O}^{.}$ through the interaction of CO with surface oxygen to
form $CO_{2}$ and neutral oxygen vacancies by the defect reaction
$CO + O_{O} \rightarrow CO_{2} + V_{O}^{\times}$. This forms a
paramagnetic defect through the defect reaction $V_{O}^{\times}
\rightarrow V_{O}^{.} + e^{'}$, producing an electron in the
conduction band of Sn 5s character as evidenced in EPR by a
resonance at g = 1.99 assigned to an unpaired electron on tin
\cite{ref43}. $V_{O}^{.}$ is thought to transfer electrons onto tin
ions to form $V_{O}^{..}$ through the reaction $Sn_{Sn}^{\times} +
2V_{O}^{.} \rightarrow Sn_{Sn}^{''} + 2V_{O}^{..}$. Shallow donor
levels for $\frac {V_{O}^{.}}{V_{O}^{\times}}$ and $\frac
{V_{O}^{..}}{V_{O}^{.}}$ have been identified 0.03 and 0.15 eV below
the conduction band minimum (CBM), respectively (see figure 2)
[38,44].\\\\As these states lie quite close to the CBM,
they will not cause a loss of transparency but will enhance the
conductivity by introducing carrier electrons into the conduction
band. Further evidence for the existence of these shallow states
comes from cathodoluminescence studies which identified a band at
1.94 eV to result from a transition between a surface oxygen vacancy
level at 1.4 eV above the valence band maximum (VBM) and a bulk
shallow donor level at 0.15 eV below the CBM \cite{ref44}.\\\\In this paper we report the structural, microstructural and optical properties of the bulk
$Sn_{1-x}Sb_{x}O_{2-\delta}$ samples with x = 0.00, 0.10, 0.20, 0.30
synthesized by solid-state reaction route.
%%%%%%%%%%%%%%%%%%%%%%%%%%%%%%%%%%%%%%%%%%%%%%%%%%%%%%%%%%%%%%%%%
\section{Experimental details}
In the present investigation firstly the bulk $SnO_{2}$ was
synthesized by oxidizing the fine powder (50 mesh) of metallic tin
(99.99\% Aldrich chemical USA) at $700^{0}C$ for 8 hrs in
programmable temperature controlled SiC furnace. Thermal oxidation
of metallic tin powder can be expressed simply by the following
reactions:\\
%%%%%%%%%%%%%%%%%%%%%%%%%%%%%%%%%%%%%%%%%%%%%%%%%%%%%%%%%%%%%%%%%%%%%%%%
\begin{equation}
\label{eq1} Sn_{(s)} + \frac{1}{2}O_{2(g)} \rightarrow SnO_{(v)},
\end{equation}
%%%%%%%%%%%%%%%%%%%%%%%%%%%%%%%%%%%%%%%%%%%%%%%%%%%%%%%%%%%%%%%%%%%%%%%%%
%%%%%%%%%%%%%%%%%%%%%%%%%%%%%%%%%%%%%%%%%%%%%%%%%%%%%%%%%%%%%%%%%%%%%%%%
\begin{equation}
\label{eq2} SnO_{(v)} + \frac{1}{2}O_{2(g)} \rightarrow SnO_{2(s)},
\end{equation}
%%%%%%%%%%%%%%%%%%%%%%%%%%%%%%%%%%%%%%%%%%%%%%%%%%%%%%%%%%%%%%%%%%%%%%%%%
%%%%%%%%%%%%%%%%%%%%%%%%%%%%%%%%%%%%%%%%%%%%%%%%%%%%%%%%%%%%%%%%%%%%%
In second step the as synthesized $SnO_{2}$ powder was converted
into form of pellets. For this we employed the most widely used
technique i.e. dry pressing, which consists of filling a die with
powder and pressing at $400 \frac{Kg}{Cm^{2}}$ into a compacted disc
shape. In this way several cylindrical pellets of 2 mm thickness and
10 mm in diameter were prepared. Finally these pellets of $SnO_{2}$
were put into alumina crucibles and sintered at $900^{0}C$ in air
for 8 hrs.\\\\Similarly samples with nominal composition
$Sn_{1-x}Sb_{x}$ $O_{2-\delta}$ (where x = 0.10, 0.20, 0.30) were
synthesized by standard solid-state reaction method. The appropriate
ratio of the constituent oxides i.e. $SnO_{2}$ (as synthesized) and
$Sb_{2}O_{3}$ (99.99\% Aldrich Chemical USA) were thoroughly mixed
and ground for several hours (2 to 4 hrs) with the help of mortar
and pestle. The mixed powder was pressed into cylindrical pellets of
2 mm thickness and 10 mm in diameter using PVA (polyvinyl alcohol)
as a binder at a pressure of $400
\frac{Kg}{Cm^{2}}$. Finally the antimony doped tin
oxide pellets were sintered in a programmable SiC furnace at
$1300^{0}C$ for 16 hrs according to the temperature profile given in
figure 3. The sintering conditions employed for all the samples are
given in table 1.
%%%%%%%%%%%%%%%%%%% Figure 3 %%%%%%%%%%%%%%%%%%%%%%%%%%%%%%%%%%%
\begin{figure}[h]
  \begin{center}
    \includegraphics[width=8.5cm,height=7cm,angle=0]{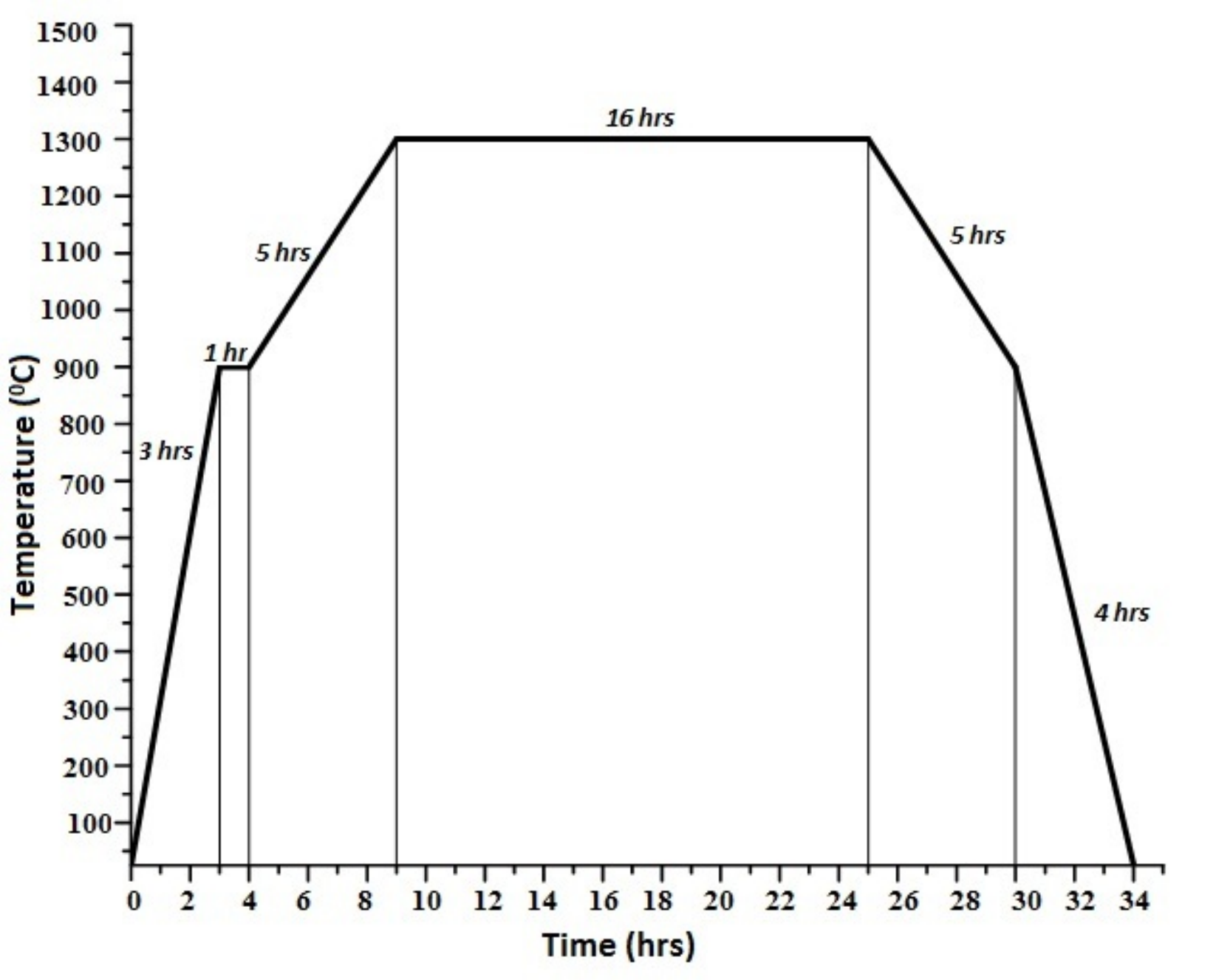}
  \end{center}

  \caption{\small The heat treatment schedule for Sb-doped
$SnO_{2}$.}
  \label{fig-label}
\end{figure}
%%%%%%%%%%%%%%%%%%%%%%%%%%%%%%%%%%%%%%%%%%%%%%%%%%%%%%%%%%%%%%%%%
%%%%%%%%%%%%%%%%%%%%%%%%%%%%%%%%%%%%%%%%%%%%%%%%%%%%%%%%%%%%%%%%%%%%%%%%%%%%%%%
\begin{table}
{\bf {Table-1. Nominal composition and sintering condition are shown for typical samples.}} \\
\mbox{ }
%\begin{ruledtabular}
\begin{tabular}{lccccc}
\\ \hline \hline \\
%\hline
{\bf Nominal} & {\bf Composition}& {\bf Sintering} \\
{\bf composition} &   & {\bf conditions} \\ \hline \\
x = 0.00  & $SnO_{2}$  &$900^{0}C$ for 8 hrs  \\
& & in air\\

x = 0.10  & $Sn_{0.90}Sb_{0.10}O_{2-\delta}$  &$1300^{0}C$ for 16 hrs\\
&  &  in air (see fig. 3)\\

x = 0.20 & $Sn_{0.80}Sb_{0.20}O_{2-\delta}$ & $1300^{0}C$ for 16 hrs\\
&  &  in air (see fig. 3)\\

x = 0.30 & $Sn_{0.70}Sb_{0.30}O_{2-\delta}$ &$1300^{0}C$ for 16 hrs\\
&  &  in air (see fig. 3)\\
\hline \hline
\end{tabular}
\end{table}The gross structure and phase purity of the powder
sample were examined by X-ray diffraction technique. The X-ray
diffraction pattern of the compounds was recorded at room
temperature using X-ray powder diffractometer (18 KW Rigaku, Japan)
with $CuK_{\alpha}$ radiation $(\lambda = 1.5418 {\AA})$ in a wide
range of Bragg angles $2\theta (20^{o}< 2\theta <80^{o})$ with
scanning rate of $2^{o} min^{-1}$. The surface morphology of as
synthesized materials has been carried out by a Joel scanning
electron microscope (JSM-5600) operated at 25 KV, with a resolution
of 3.5 nm. The structural/microstructural characteristics was
explored by transmission electron microscope (TEM, FEI Tecnai $G^{2}
20$) in both the imaging and diffraction modes. Optical studies were
performed by measuring the reflectance and the absorbance in the
wavelength region 200 - 800 nm at room temperature using a Varian
Cary 5000 UV-Vis spectrophotometer.
\section{Results and discussion}
Figure 4 shows the powder x-ray diffraction patterns of
$Sn_{1-x}Sb_{x}O_{2-\delta}$ with x = 0.00, 0.10, 0.20, 0.30.
%%%%%%%%%%%%%%%%%%% Figure 4 %%%%%%%%%%%%%%%%%%%%%%%%%%%%%%%%%%%
\begin{figure}[h]
  \begin{center}
    \includegraphics[width=8.5cm,height=7.5cm,angle=0]{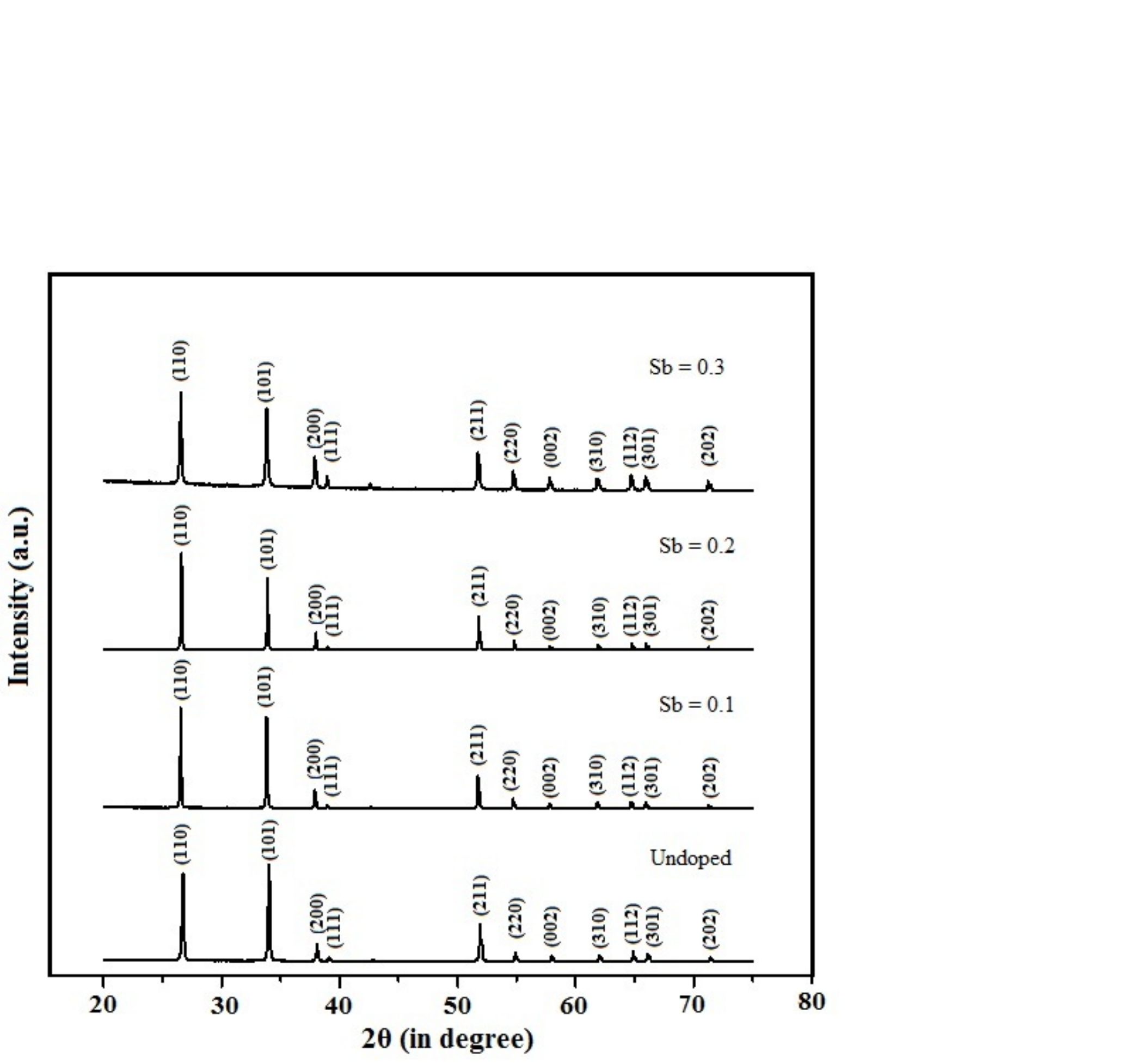}
  \end{center}

  \caption{\small X-ray powder diffraction patterns of
$Sn_{1-x}Sb_{x}O_{2-\delta}$ system with x = 0.00, 0.10, 0.20,
0.30.}
  \label{fig-label}
\end{figure}
The analysis of x-ray diffraction patterns revealed
that the as synthesized doped and undoped tin oxides are pure
crystalline tetragonal rutile phase of tin oxide (JCPDS card no.
041-1445) which belongs to the space group $P4_{2}/mnm$ (number
136). No obvious reflection peaks from impurities, such as unreacted
Sn, Sb or other oxide phases such as $Sb_{2}O_{5}$ or $Sb_{2}O_{3}$
are detected, indicating high purity of the product. It is
perceptible from the XRD lines of figure 4 that the undoped as well
as doped tin oxide samples grow along the preferred orientation of
(110) and (101) reflections. The presence of other reflections such
as (200), (111), (211), (220), (002), (310), (112), (301) and (202)
have also been detected with considerable intensities for both doped
and undoped tin oxide samples.\\\\We have calculated the
lattice parameters using high angle XRD lines such as (002), (310),
(112), (301) and (202) shown in figure 4. The calculated lattice
parameters of $Sn_{1-x}Sb_{x}O_{2-\delta}$ (x = 0.00, 0.10, 0.20,
0.30) are shown in Table 2.
%%%%%%%%%%%%%%%%%%%%%%%%%%%%%%%%%%%%%%%%%%%%%%%%%%%%%%%%%%%%%%%%%%%%%%%%%%%%%%%
\begin{table}
{\bf {Table-2. Lattice parameters of $Sn_{1-x}Sb_{x}O_{2-\delta}$ (x
= 0.00, 0.10, 0.20,
0.30)}} \\
\mbox{ }
%\begin{ruledtabular}
\begin{tabular}{lccccc}
\\ \hline \hline \\
%\hline
&{\bf Nominal} & {\bf Lattice parameters}& &\\
&{\bf Composition of Sb} & {\bf a = b $({\AA})$}  & {\bf c $({\AA})$ } &\\ \hline \\
&x = 0.00  & 4.7346  &3.1787 \\

&x = 0.10  & 4.7354  &3.1822 \\

&x = 0.20 & 4.7360 &3.1843\\

&x = 0.30 & 4.7365 &3.1899\\

\hline \hline 
\end{tabular}
\end{table}
%%%%%%%%%%%%%%%%%%%%%%%%%%%%%%%%%%%%%%%%%%%%%%%%%%%%%%%%%%%%
A small increase in
the lattice parameters of the tetragonal unit cell has been observed
with increasing Sb content (figure 5 \& 6).\\\\
\begin{table}
{\bf {Table-3. Crystallite sizes determined by XRD of undoped tin oxide sample}} \\
\mbox{ }
%\begin{ruledtabular}
\begin{tabular}{lccccc}
\\ \hline \hline \\
%%\hline
{\bf Diffraction line} & {\bf FWHM $\beta$}& {\bf
Crystallite size}\\
& {\bf(degree)}& {\bf (nm)}\\ \hline \\
110  & 0.1248  &65.4 \\

101  & 0.1440  &57.7 \\

200 & 0.1056 &79.6\\

111 & 0.0576 &146.4\\

211 & 0.1152 &76.7\\

220 & 0.1056  &84.8\\

002 & 0.0864 &105.1\\

310 & 0.1056  &87.7\\

112 & 0.1056 &89.1\\

301 & 0.1248  &75.9\\

202 & 0.0864  &113.1\\
\hline \hline
\end{tabular}
\end{table}
%%%%%%%%%%%%%%%%%%%%%%%%%%%%%%%%%%%%%%%%%%%%%%%%%%%%%%%%%%%%%%%%%%%%%%%%%%%%%%%%%%%%%%%%
%%%%%%%%%%%%%%%%%%% Figure 5 %%%%%%%%%%%%%%%%%%%%%%%%%%%%%%%%%%%
\begin{figure}[t]
  \begin{center}
    \includegraphics[width=8cm,height=7cm,angle=0]{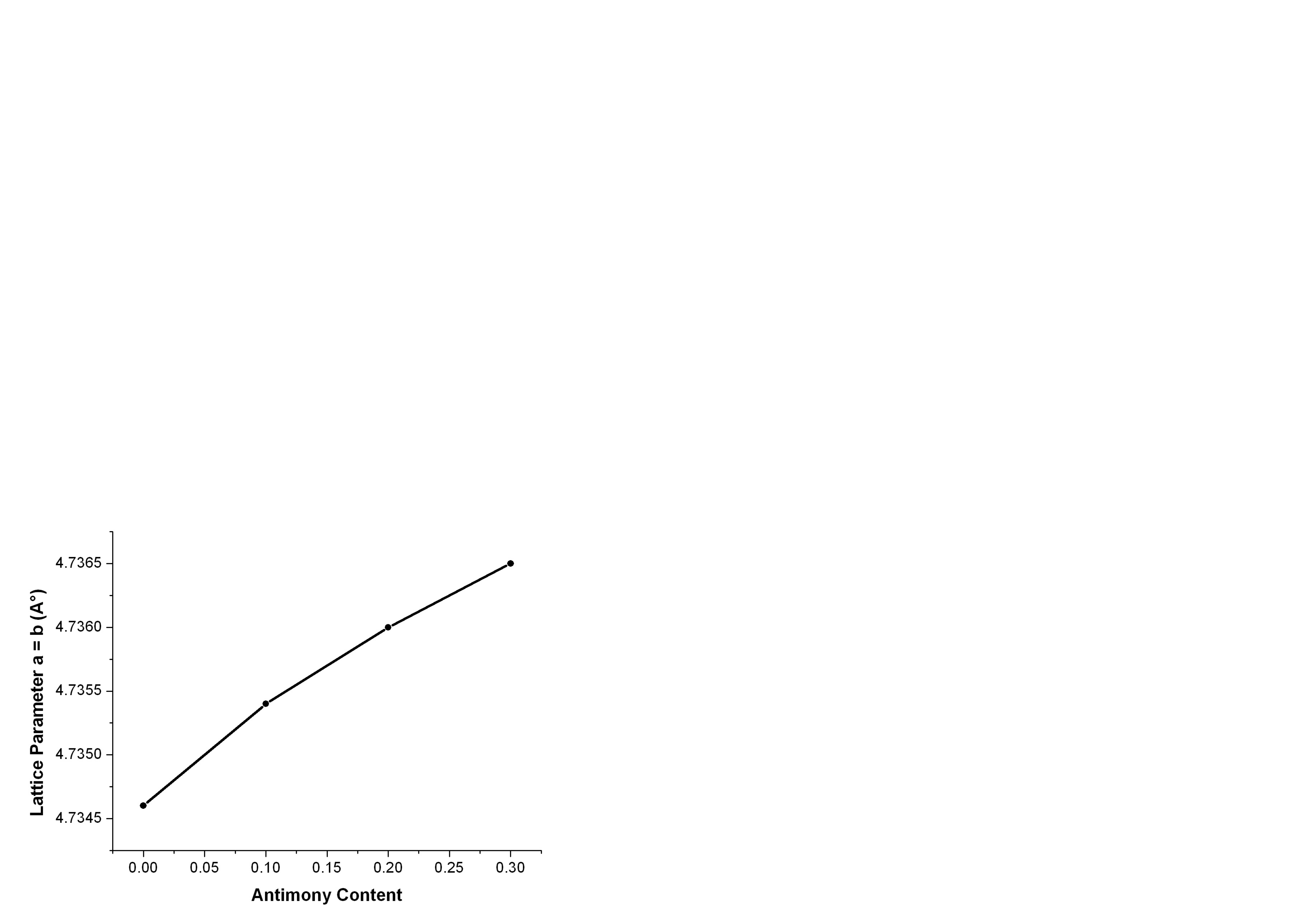}
  \end{center}

  \caption{\small The variation of lattice parameter `a' (or `b')
versus Sb - content shows minute elongation in lattice parameter `a'
(or `b').}
  \label{fig-label}
\end{figure}
%%%%%%%%%%%%%%%%%%%%%%%%%%%%%%%%%%%%%%%%%%%%%%%%%%%%%%%%%%%%%%%%%
%%%%%%%%%%%%%%%%%%% Figure 6 %%%%%%%%%%%%%%%%%%%%%%%%%%%%%%%%%%%
\begin{figure}[t]
  \begin{center}
    \includegraphics[width=8cm,height=7cm,angle=0]{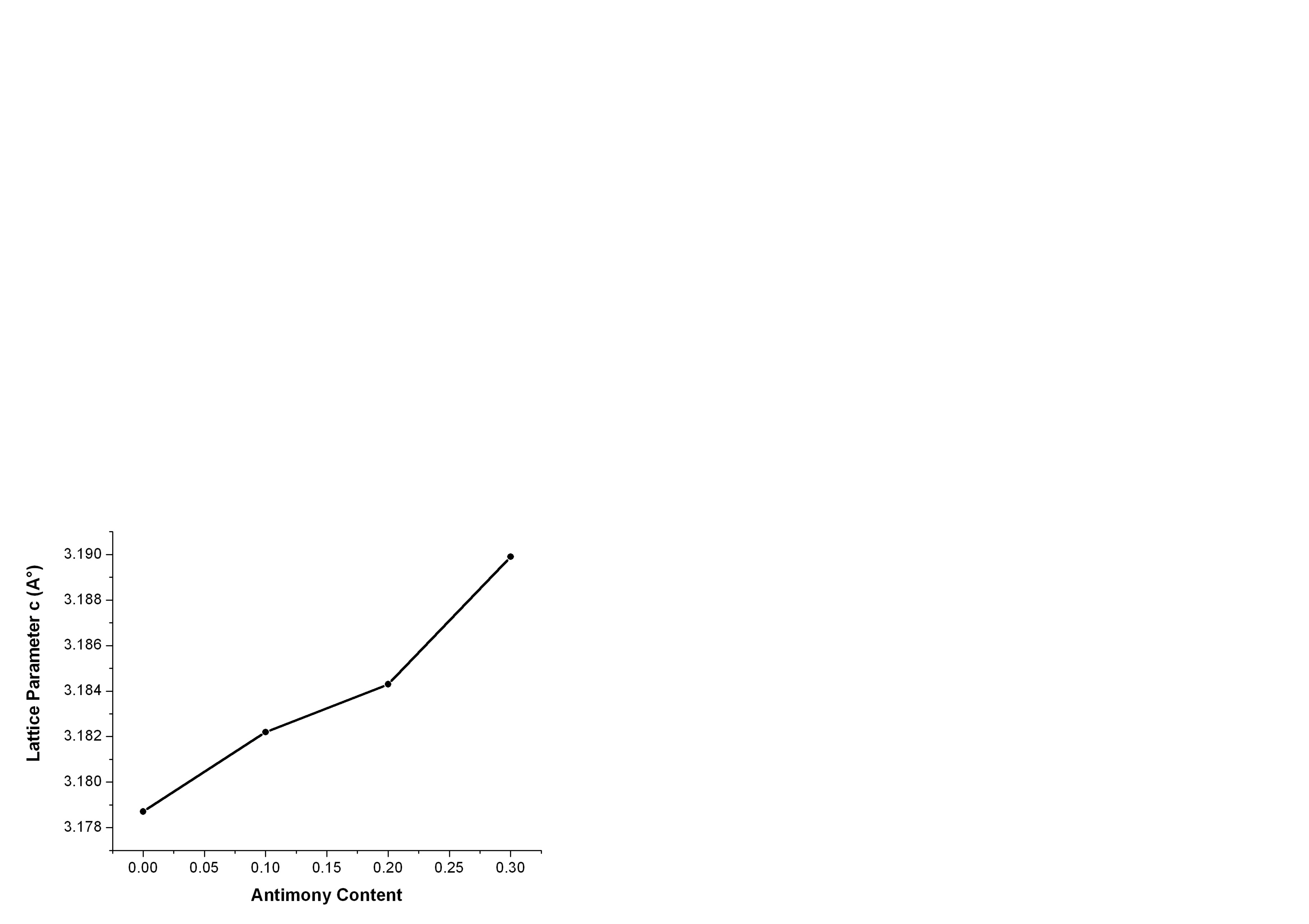}
  \end{center}

  \caption{\small The variation of lattice parameter `c' versus Sb -
content shows minute elongation in lattice parameter `c'.}
  \label{fig-label}
\end{figure}
%%%%%%%%%%%%%%%%%%%%%%%%%%%%%%%%%%%%%%%%%%%%%%%%%%%%%%%%%%%%%%%%%
For example for x = 0.00, $\textbf{a} = 4.7346{\AA}$, $\textbf{c} =
3.1787{\AA}$ whereas for x = 0.30, $\textbf{a} = 4.7365{\AA}$,
$\textbf{c} = 3.1899{\AA}$. This may possibly occur due to the
difference in ionic radii of $Sn^{4+} (0.72{\AA})$ and $Sb^{3+}
(0.90{\AA})$ ions. The diffraction peaks are markedly broadened, which
indicates that the crystalline sizes of samples are small.
Crystallite size was
automatically calculated from x-ray diffraction data using the Debye-Scherrer formula, \cite{ref45}:\\
%%%%%%%%%%%%%%%%%%%%%%%%%%%%%%%%%%%%%%%%%%%%%%%%%%%%%%%%%%%%%%%%%%%%%%%%
\begin{equation}
\label{eq3}
 D_{hkl} = \frac{0.9\lambda}{\beta\cos\theta},
\end{equation}
%%%%%%%%%%%%%%%%%%%%%%%%%%%%%%%%%%%%%%%%%%%%%%%%%%%%%%%%%%%%%%%%%%%%%%%%%
where $\lambda$ is the x-ray wavelength (1.5418{\AA} for
$CuK_{\alpha}$), $\theta$ is the Bragg angle and $\beta$ is the full
width of the diffraction line at half its maximum intensity (FWHM).
The crystallite sizes were calculated (using equation 3) to be
$\sim$ 65, 57, 77 nm for the undoped tin oxide sample corresponding
to the diffraction peaks 110, 101 and 211 respectively, as shown in
Table 3. These values are in agreement with those observed by SEM
and TEM investigation as will be shown later.\\\\The scanning electron 
micrographs (SEM) of antimony
substituted $SnO_{2}$ samples are shown in figures 7 to 10. These
electron micrographs reveal that the grain size in the antimony
doped sample is larger than that of undoped one. In undoped sample
the grains are well connected (see figure 7). When 0.10 mole of
antimony is added to $SnO_{2}$ sample, few pores are found between
regions of well connected grains. The pores/voids between the grains
also increase with antimony concentration up to 0.30 as shown in
figures 8 to 10. Also we did not observe any phase aggregation other
than the increase in grain size due to Sb doping. This observation
obtained from SEM studies supports the XRD results.\\\\The transmission electron micrographs (TEM) and its corresponding
selected-area electron diffraction (SAED) patterns of antimony doped
$SnO_{2}$ samples are shown in figures 11 to 18. TEM image of
undoped sample (figure 11) indicates that the $SnO_{2}$ grains have
diameters ranging from 25 to 120 nm and most grains are in cubic or
spherical shaped. These grains size are quite similar to those
calculated from Scherrer's equation 3 (see table 3). The SAED
pattern shown in figure 12 taken from pure $SnO_{2}$ sample shows
several sharp rings, which were indexed to the (110), (101), (211),
and (301) planes of the rutile crystalline structure of $SnO_{2}$
(JCPDS card no. 041-1445). It is clear from the TEM micrographs
(figures 11 to 18) the grain size are increased (nano to micro) with
increasing the Sb concentration and we get the regular SAED pattern
with higher concentration of Sb element.\\\\The optical
reflectance spectra of $Sn_{1-x}Sb_{x}O_{2-\delta}$ (x = 0.00, 0.10,
0.20, 0.30) samples as a function of wavelength ranging from 200 to
800 nm is shown in figure 19. The reflectance in all four cases (x =
0.00, 0.10, 0.20, 0.30) is small for the wavelength range 200-300 nm
whereas it shows increase in reflection above the wavelength of 300
nm. The reflectance is found to increase with the increase in doping
concentration.\\\\Optical absorbance Spectra of
$Sn_{1-x}Sb_{x}O_{2-\delta}$ samples with different Sb
concentrations (x = 0.00, 0.10, 0.20, 0.30) is shown in figure 20.
It is apparent from this figure that the absorbance increases for
the wavelength range 200 - 310 nm whereas it decreases for the
wavelength range 365 - 800 nm with the increase in doping
concentration.\\\\The transmittance of the as synthesized samples was
calculated from reflectance (figure 19) and  absorbance (figure 20)
spectra using the relation:
%%%%%%%%%%%%%%%%%%%%%%%%%%%%%%%%%%%%%%%%%%%%%%%%%%%%%%%%%%%%%%%%%%%%%%%%
\begin{equation}
\label{eq4}
 T = 1 - (R + A),
\end{equation}
%%%%%%%%%%%%%%%%%%%%%%%%%%%%%%%%%%%%%%%%%%%%%%%%%%%%%%%%%%%%%%%%%%%%%%%%%
where,\\\\ T = Transmittance\\R = Reflectance\\A = Absorbance.\\\\The transmittance spectra calculated (using equation 4)
for the bulk $Sn_{1-x}-$ $Sb_{x}O_{2-\delta}$ (x = 0.00, 0.10, 0.20,
0.30) samples in the wavelength range of 200 - 800 nm is shown in
figure 21. The figure clearly shows the increase in transmittance
due to increase in antimony (Sb) concentration for the wavelength
range 360 - 800 nm whereas it shows the decrease in transmittance at
the wavelength below $\sim$ 305 nm.\\\\The
variation of the optical absorption coefficient $\alpha$ with photon
energy $h\nu$ was obtained using the absorbance data (figure 20) for
various samples. The absorption coefficient $\alpha$ may be written
as a function of the incident photon energy $h\nu$ \cite{ref46}:
%%%%%%%%%%%%%%%%%%%%%%%%%%%%%%%%%%%%%%%%%%%%%%%%%%%%%%%%%%%%%%%%%%%%%%%%
\begin{equation}
\label{eq5}
 \alpha = \frac{A}{h\nu}(h\nu - E_{g})^{n},
\end{equation}
%%%%%%%%%%%%%%%%%%%%%%%%%%%%%%%%%%%%%%%%%%%%%%%%%%%%%%%%%%%%%%%%%%%%%%
where A is a constant which is different for different transitions
indicated by different values of n, and $E_{g}$ is the corresponding
bandgap. For direct transitions $ n = \frac{1}{2}$ or $n =
\frac{2}{3}$, while for indirect ones n = 2 or 3, depending on
whether they are allowed or forbidden, respectively \cite{ref46}.
Many groups have used the above formula to calculate the bandgap of
$SnO_{2}$ samples and reported that $SnO_{2}$ is a direct bandgap
material [47-51]. The bandgap can be deduced from a plot of $(\alpha
h\nu)^{2}$ versus photon energy $(h\nu)$. Better linearity of these
plots suggests that the samples have direct band transition. The
extrapolation of the linear portion of the $(\alpha h\nu)^{2}$ vs.
$h\nu$ plot to $\alpha = 0$ will give the bandgap value of the
samples \cite{ref52}.\\\\
Figures 22 to 25
shows the $(\alpha h\nu)^{2}$ versus photon
energy $(h\nu)$ plot for pure and antimony (Sb) doped $SnO_{2}$
samples. The linear fits obtained for these plots are also depicted
in the figures. The bandgap $(E_{g})$ values for the
$Sn_{1-x}Sb_{x}O_{2-\delta}$ samples with antimony (Sb)
concentrations x = 0.00, 0.10, 0.20 and 0.30 are 3.367 eV, 3.406 eV,
3.511 eV and 3.558 eV respectively (as shown in figures 22 to 25).
From figures 22 to 25, it is observed that the
bandgap of the samples increases with the increase in concentration
of antimony (Sb). This is consistent with that
reported in [53,54]. The bandgap is plotted as a function of
increasing Sb concentration in figure 26, and also the $E_{g}$
values are given in table 4.
The increased of transmittance and
bandgap of the as synthesized antimony doped tin oxide samples may
be explained in terms of the enhancement of oxygen vacancies. Since
the oxygen vacancies increase due to the partial substitution of
$Sn^{4+}$ by $Sb^{3+}$.
%%%%%%%%%%%%%%%%%%%%%%%%%%%%%%%%%%%%%%%%%%%%%%%%%%%%%%%%%%%%%%%%%%%%%%%%%%%%%%%
\begin{table}
{\bf {Table-4. Grains size range and optical bandgap of the Sb doped
$SnO_{2}$ samples. Both grain size and optical bandgap
 values increases upon increasing antimony (Sb) concentration in the in the
compound ($Sn_{1-x}Sb_{x}O_{2-\delta}$).}} \mbox{ }
%\begin{ruledtabular}
\begin{tabular}{ccccc}
\\ \hline \hline\\
%\hline
{\bf Nominal} & {\bf Bandgap ($E_{g}$ (eV))} & {\bf Grains Size (Range)} \\
{\bf Composition} & & \\\hline\\
x = 0.00  & 3.367  & Nano Range \\

x = 0.10  & 3.406 & Micro Range \\

x = 0.20 & 3.511 & Micro Range\\

x = 0.30 & 3.558 & Micro Range\\

\hline \hline
\end{tabular}
\end{table}
%%%%%%%%%%%%%%%%%%%%%%%%%%%%%%%%%%%%%%%%%%%%%%%%%%%%%%%%%%%%

%%%%%%%%%%%%%%%%%%% Figure 7 %%%%%%%%%%%%%%%%%%%%%%%%%%%%%%%%%%%
\begin{figure}[htbp]
  \begin{center}
    \includegraphics[width=8cm,height=5.8cm,angle=0]{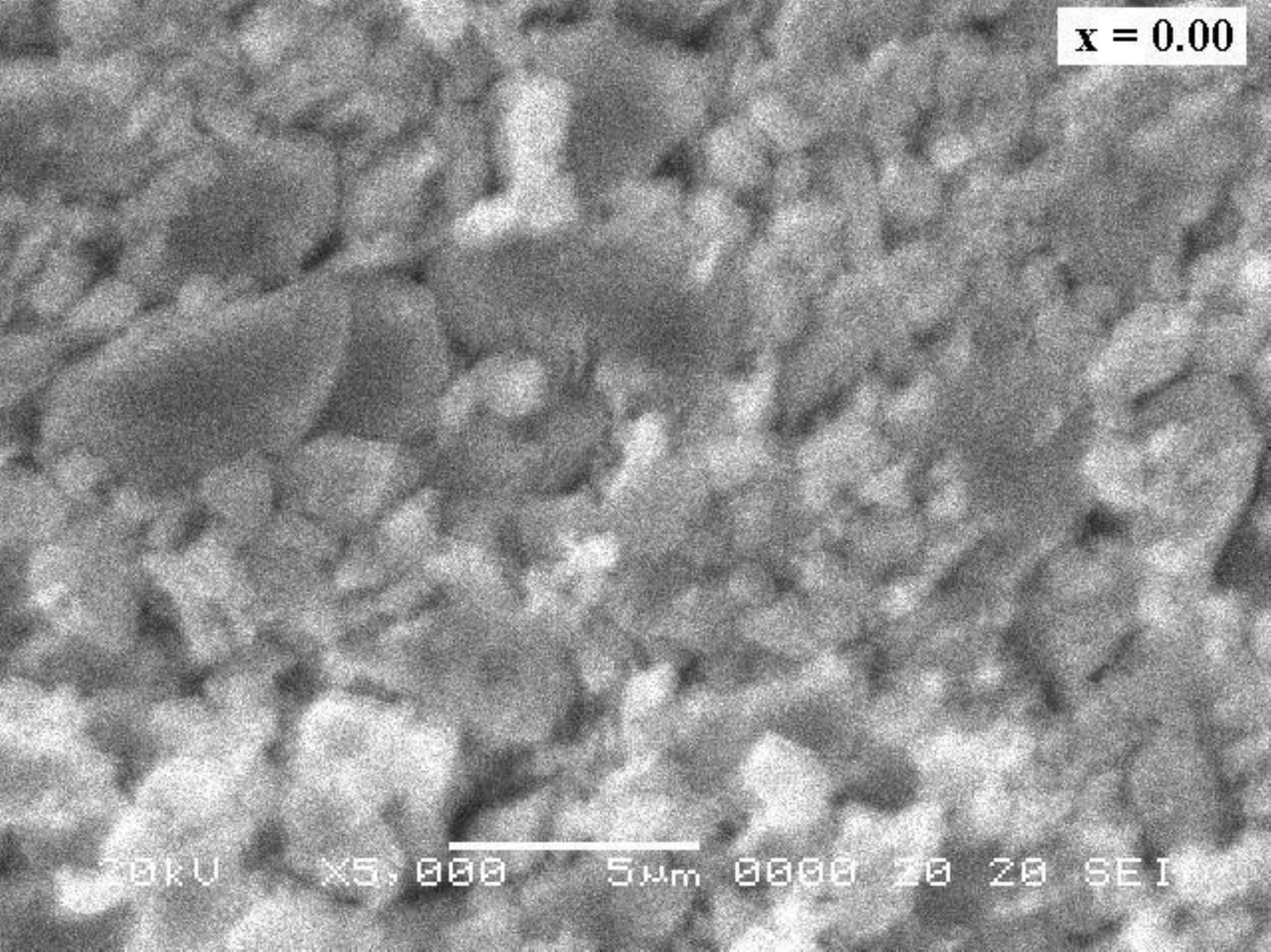}
  \end{center}

  \caption{\small Scanning electron micrograph of the surface of the
bulk $Sn_{1-x}Sb_{x}O_{2-\delta}$ system with x = 0.00.}
  \label{fig-label}
\end{figure}
%%%%%%%%%%%%%%%%%%% Figure 8 %%%%%%%%%%%%%%%%%%%%%%%%%%%%%%%%%%%
\begin{figure}[htbp]
  \begin{center}
    \includegraphics[width=8cm,height=5.8cm,angle=0]{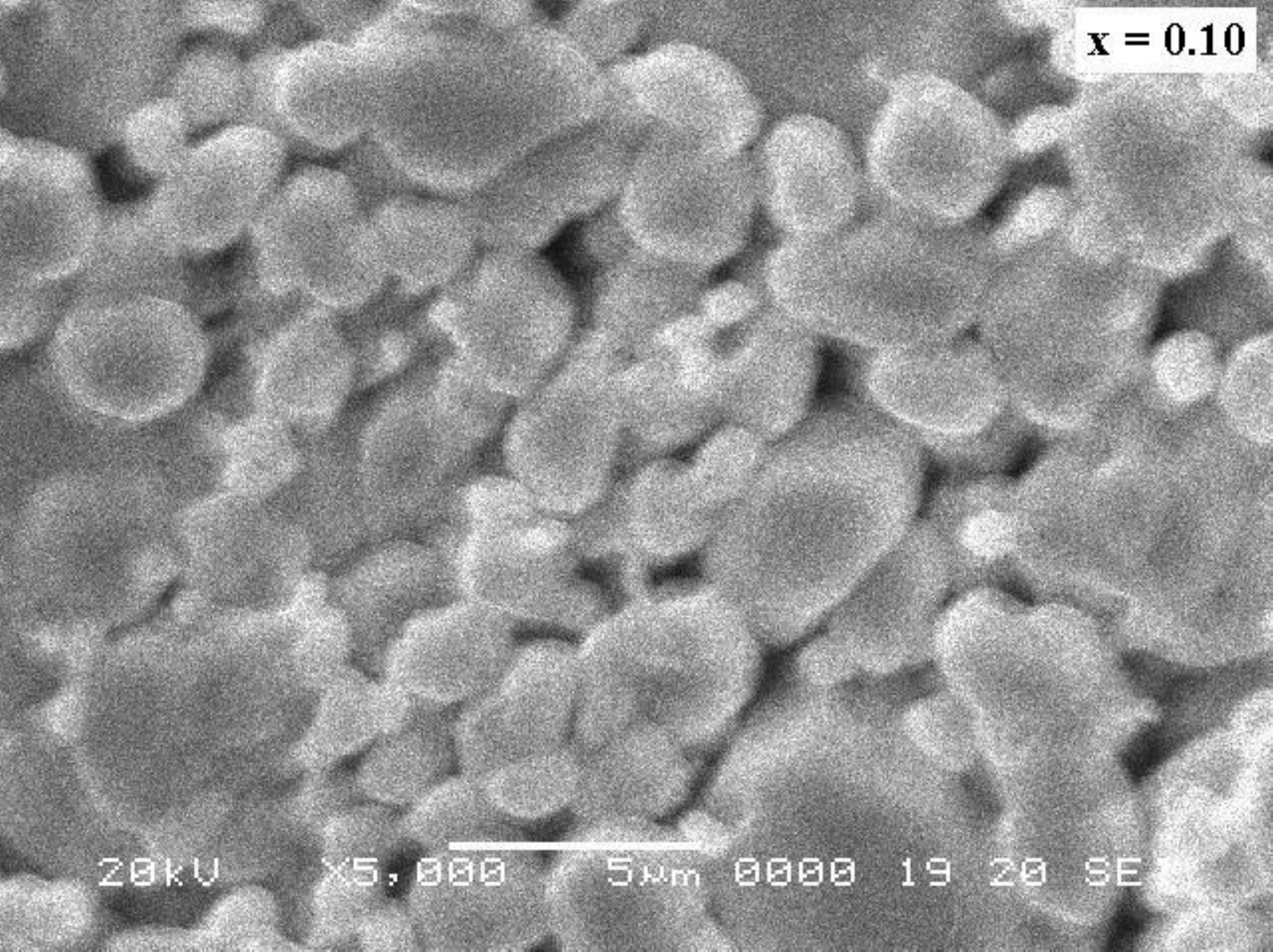}
  \end{center}

  \caption{\small Scanning electron micrograph of the surface of the
bulk $Sn_{1-x}Sb_{x}O_{2-\delta}$ system with x = 0.10.}
  \label{fig-label}
\end{figure}
%%%%%%%%%%%%%%%%%%% Figure 9 %%%%%%%%%%%%%%%%%%%%%%%%%%%%%%%%%%%
\begin{figure}[htbp]
  \begin{center}
    \includegraphics[width=8cm,height=5.8cm,angle=0]{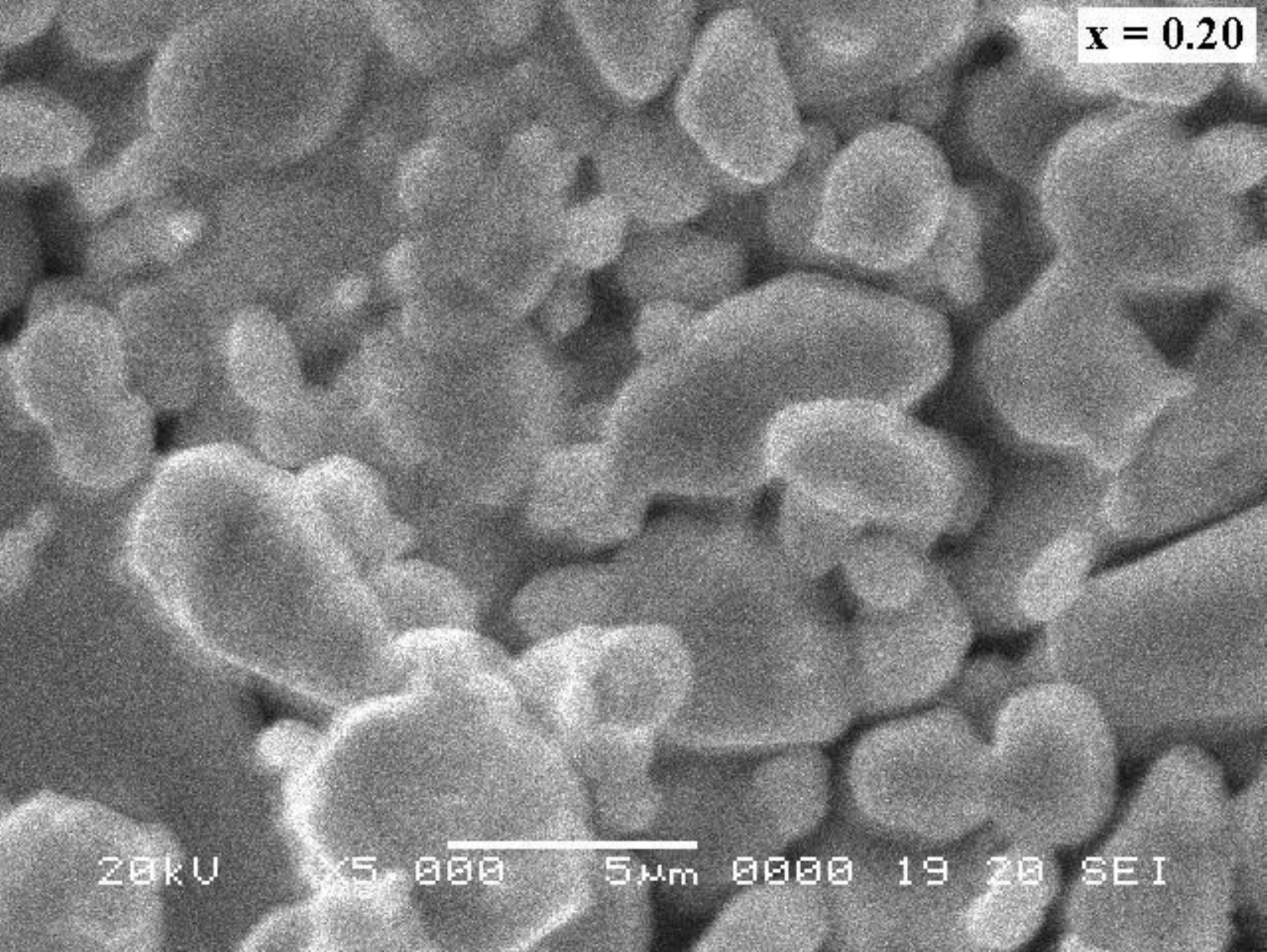}
  \end{center}

  \caption{\small Scanning electron micrograph of the surface of the
bulk $Sn_{1-x}Sb_{x}O_{2-\delta}$ system with x = 0.20.}
  \label{fig-label}
\end{figure}
%%%%%%%%%%%%%%%%%%% Figure 10 %%%%%%%%%%%%%%%%%%%%%%%%%%%%%%%%%%%
\begin{figure}[htbp]
  \begin{center}
    \includegraphics[width=8cm,height=5.8cm,angle=0]{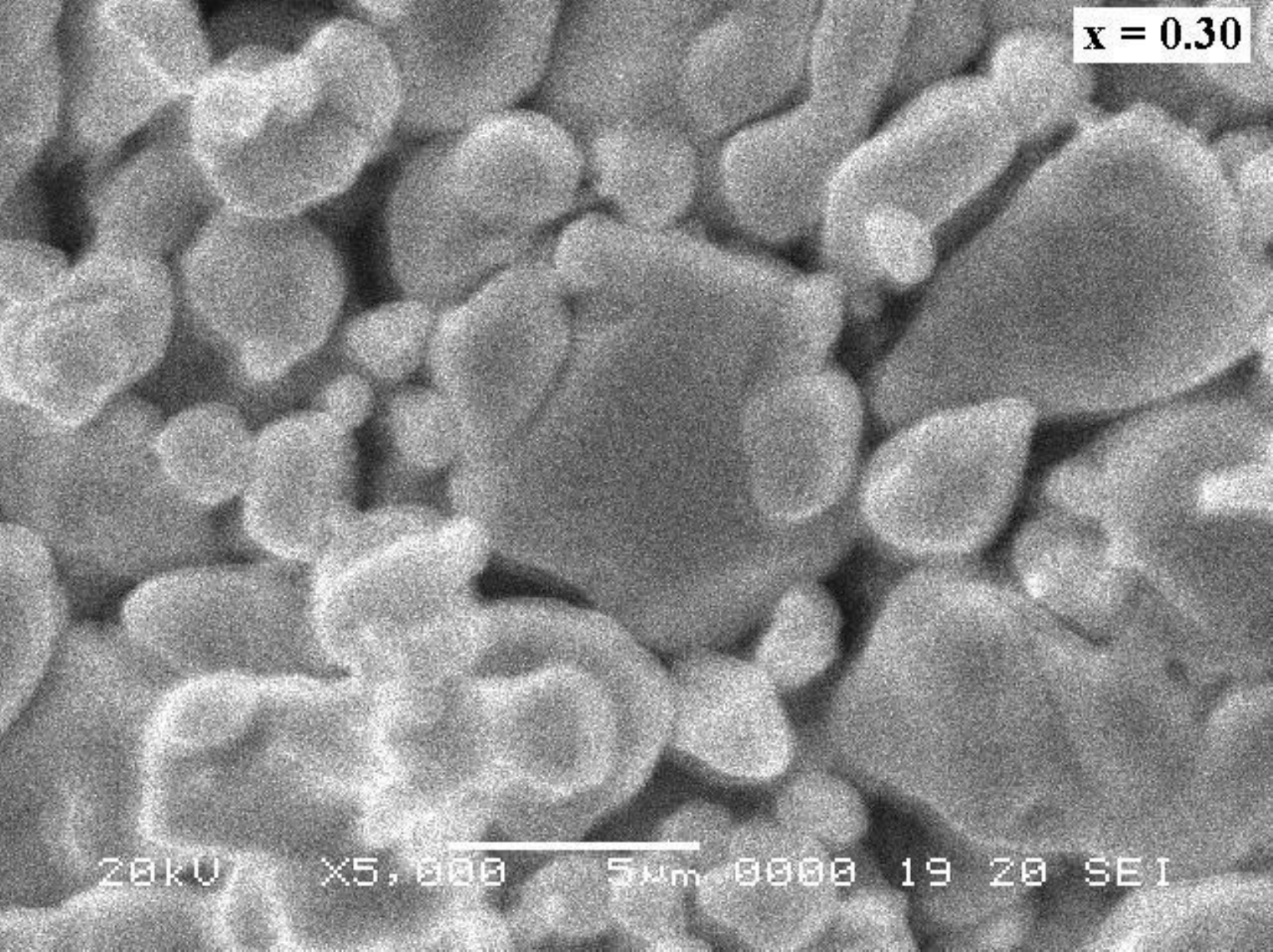}
  \end{center}

  \caption{\small Scanning electron micrograph of the surface of the
bulk $Sn_{1-x}Sb_{x}O_{2-\delta}$ system with x = 0.30.}
  \label{fig-label}
\end{figure}
%%%%%%%%%%%%%%%%%%%%%%%%%%%%%%%%%%%%%%%%%%%%%%%%%%%%%%%%%%%%%%%%%
%%%%%%%%%%%%%%%%%%% Figure 11 %%%%%%%%%%%%%%%%%%%%%%%%%%%%%%%%%%%
\begin{figure}[htbp]
  \begin{center}
    \includegraphics[width=8cm,height=5.5cm,angle=0]{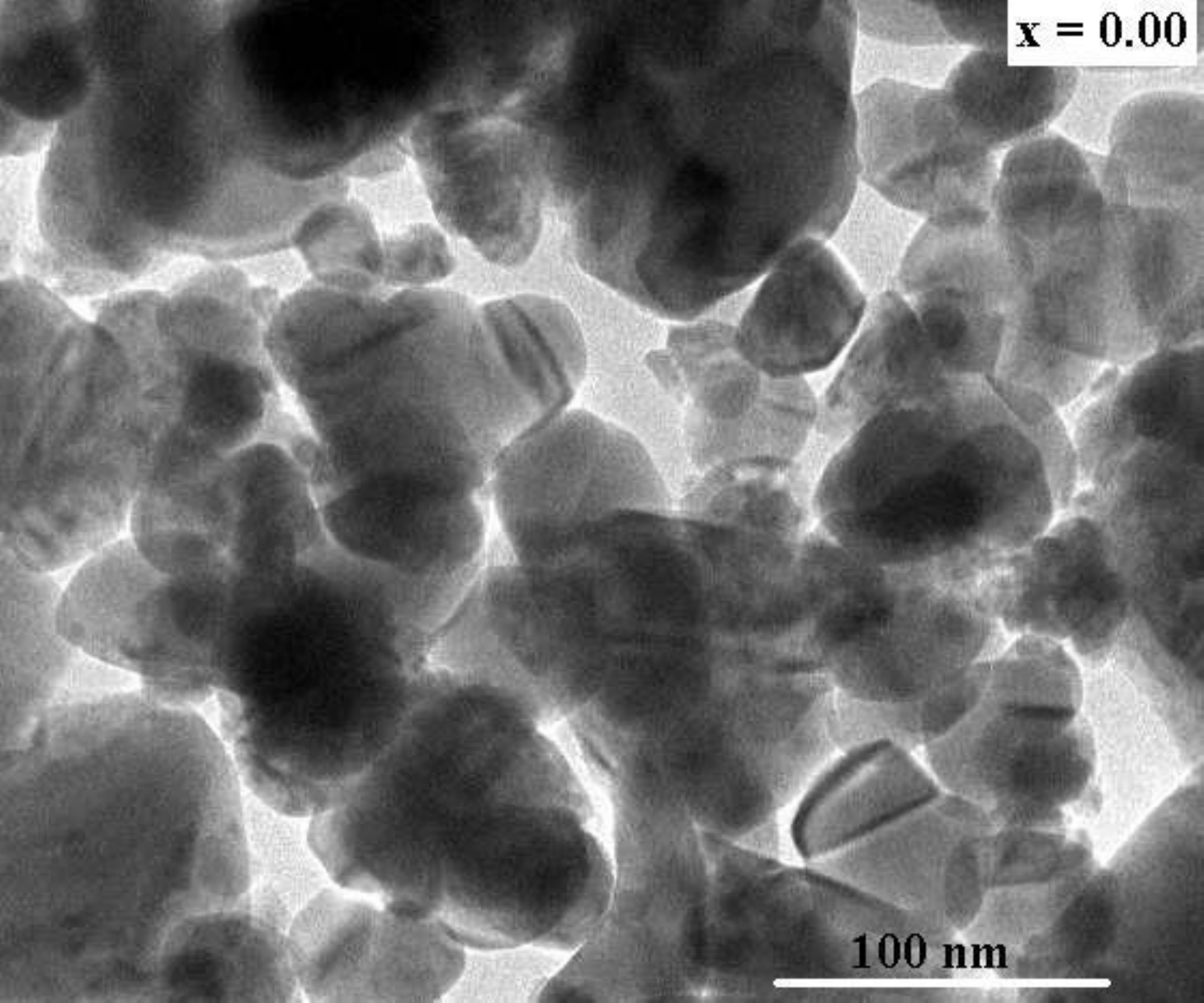}
  \end{center}

  \caption{\small Transmission electron micrograph of the bulk
$Sn_{1-x}Sb_{x}O_{2-\delta}$ system with x = 0.00 showing several
namocubes or nanospheres.}
  \label{fig-label}
\end{figure}
%%%%%%%%%%%%%%%%%%% Figure 12 %%%%%%%%%%%%%%%%%%%%%%%%%%%%%%%%%%%
\begin{figure}[htbp]
  \begin{center}
    \includegraphics[width=8cm,height=5.5cm,angle=0]{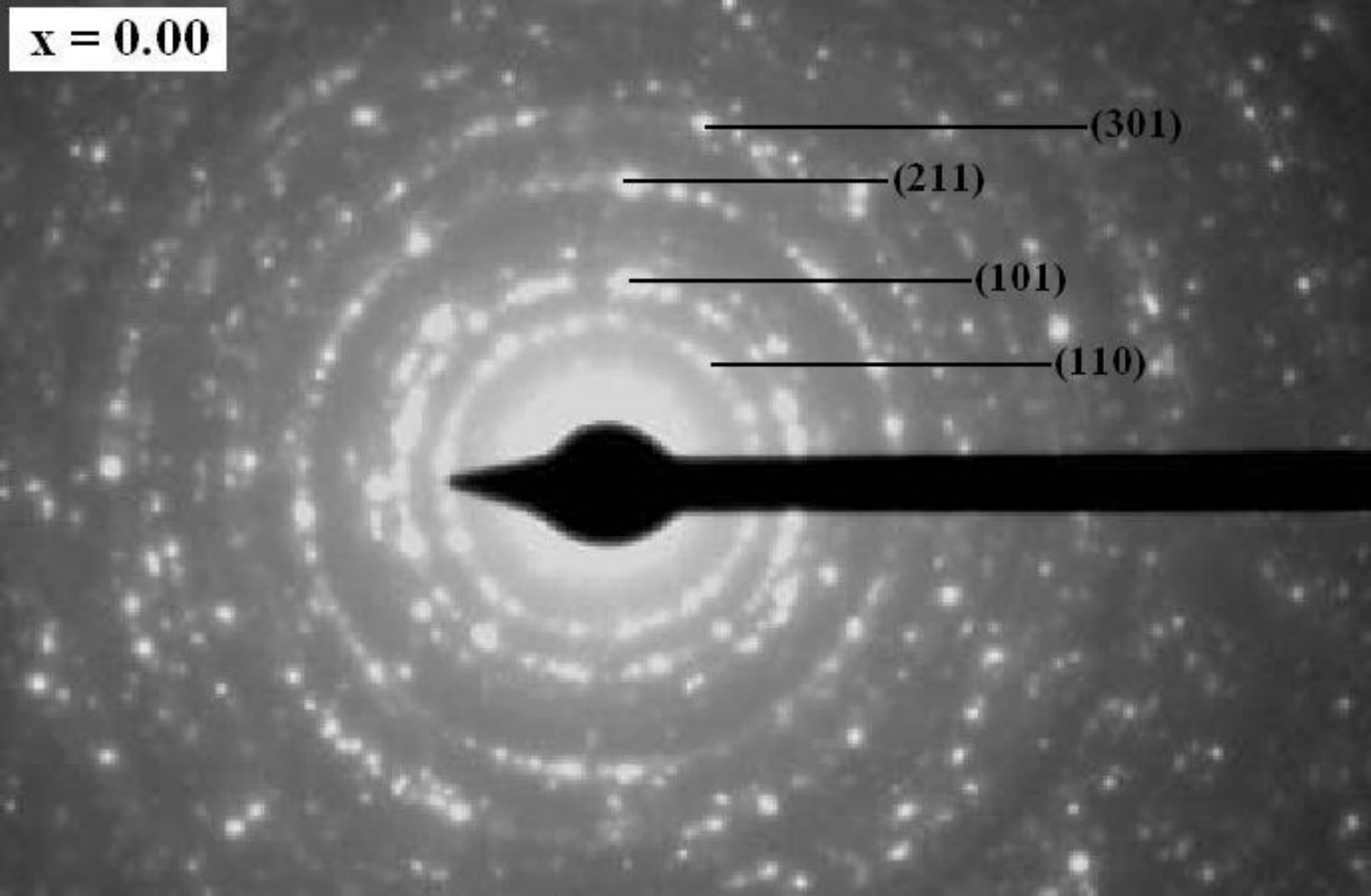}
  \end{center}

  \caption{\small Selected area electron diffraction (SAED) pattern
corresponding to figure (11) along [001] direction.}
  \label{fig-label}
\end{figure}
%%%%%%%%%%%%%%%%%%%%%%%%%%%%%%%%%%%%%%%%%%%%%%%%%%%%%%%%%%%%%%%%%
%%%%%%%%%%%%%%%%%% Figure 13 %%%%%%%%%%%%%%%%%%%%%%%%%%%%%%%%%%%
\begin{figure}[htbp]
  \begin{center}
    \includegraphics[width=8cm,height=5.5cm,angle=0]{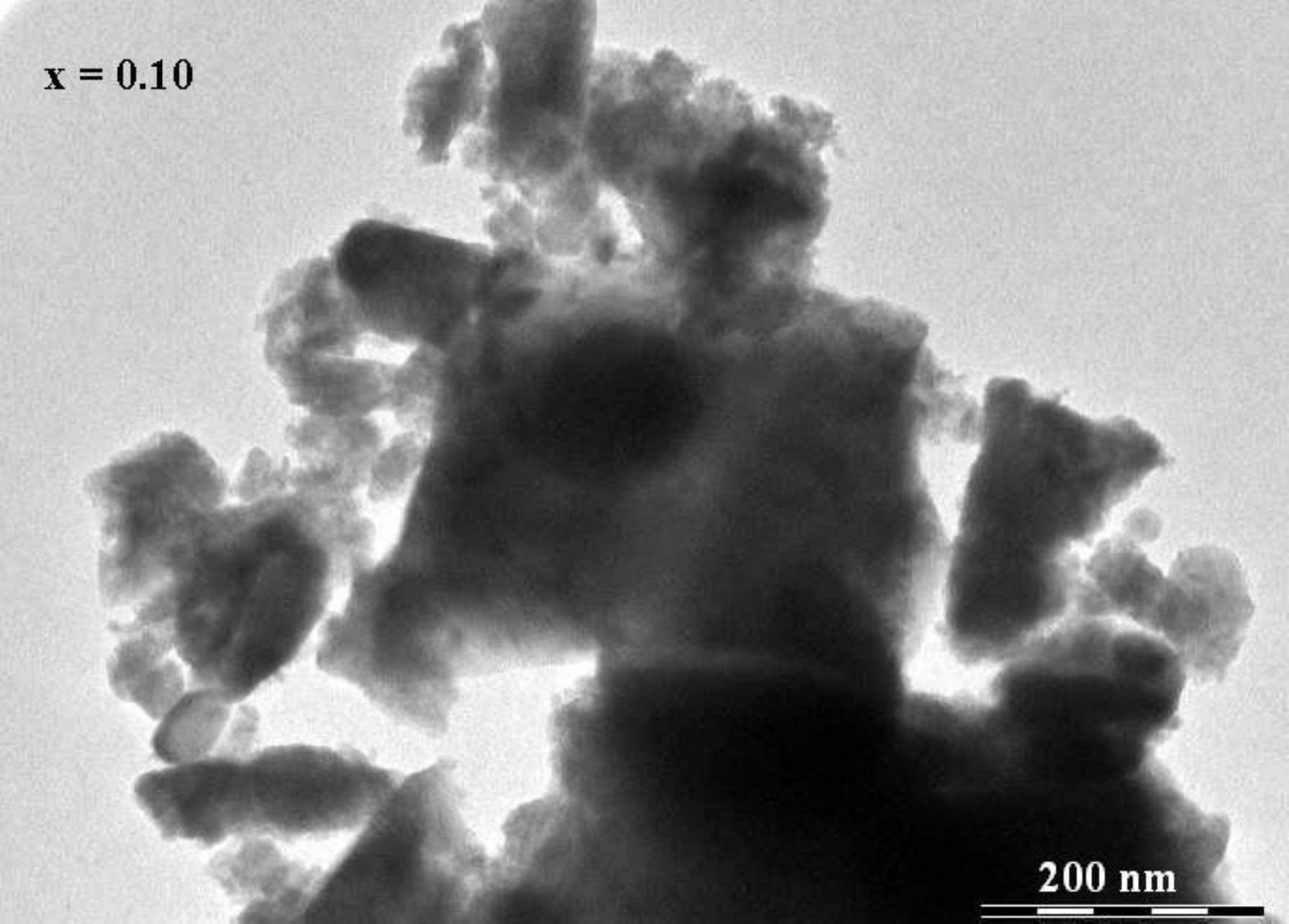}
  \end{center}

  \caption{\small Transmission electron micrograph of the bulk
$Sn_{1-x}Sb_{x}O_{2-\delta}$ system with x = 0.10 showing several
microcrystals.}
  \label{fig-label}
\end{figure}
%%%%%%%%%%%%%%%%%%% Figure 14 %%%%%%%%%%%%%%%%%%%%%%%%%%%%%%%%%%%
\begin{figure}[htbp]
  \begin{center}
    \includegraphics[width=8cm,height=5.5cm,angle=0]{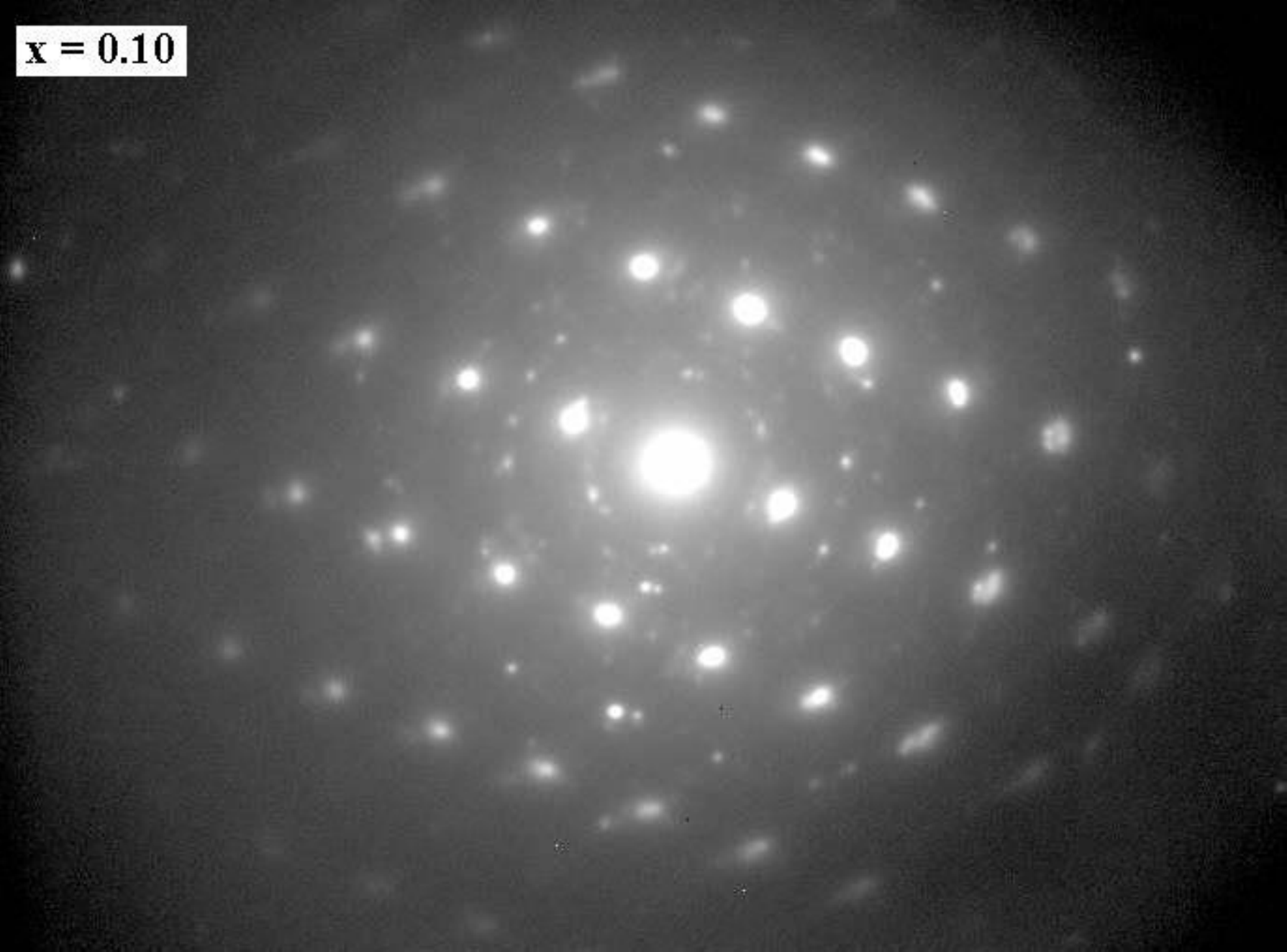}
  \end{center}

  \caption{\small Selected area electron diffraction (SAED) pattern
corresponding to figure (13) along [001] direction.}
  \label{fig-label}
\end{figure}
%%%%%%%%%%%%%%%%%%% Figure 15 %%%%%%%%%%%%%%%%%%%%%%%%%%%%%%%%%%%
\begin{figure}[htbp]
  \begin{center}
    \includegraphics[width=8cm,height=5.5cm,angle=0]{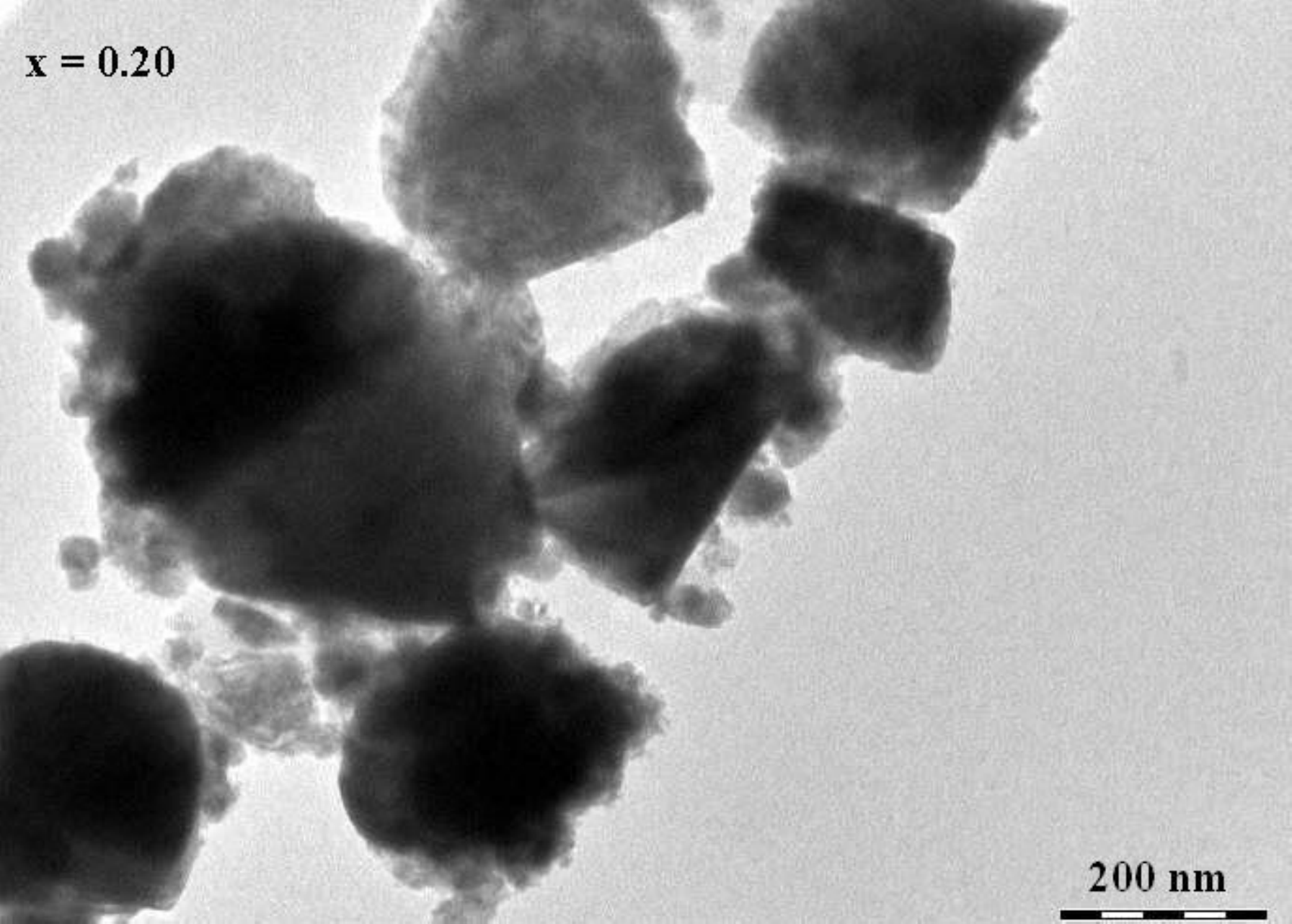}
  \end{center}

  \caption{\small Transmission electron micrograph of the bulk
$Sn_{1-x}Sb_{x}O_{2-\delta}$ system with x = 0.20 showing several
microcrystals.}
  \label{fig-label}
\end{figure}
%%%%%%%%%%%%%%%%%%% Figure 16 %%%%%%%%%%%%%%%%%%%%%%%%%%%%%%%%%%%
\begin{figure}[htbp]
  \begin{center}
    \includegraphics[width=8cm,height=5.5cm,angle=0]{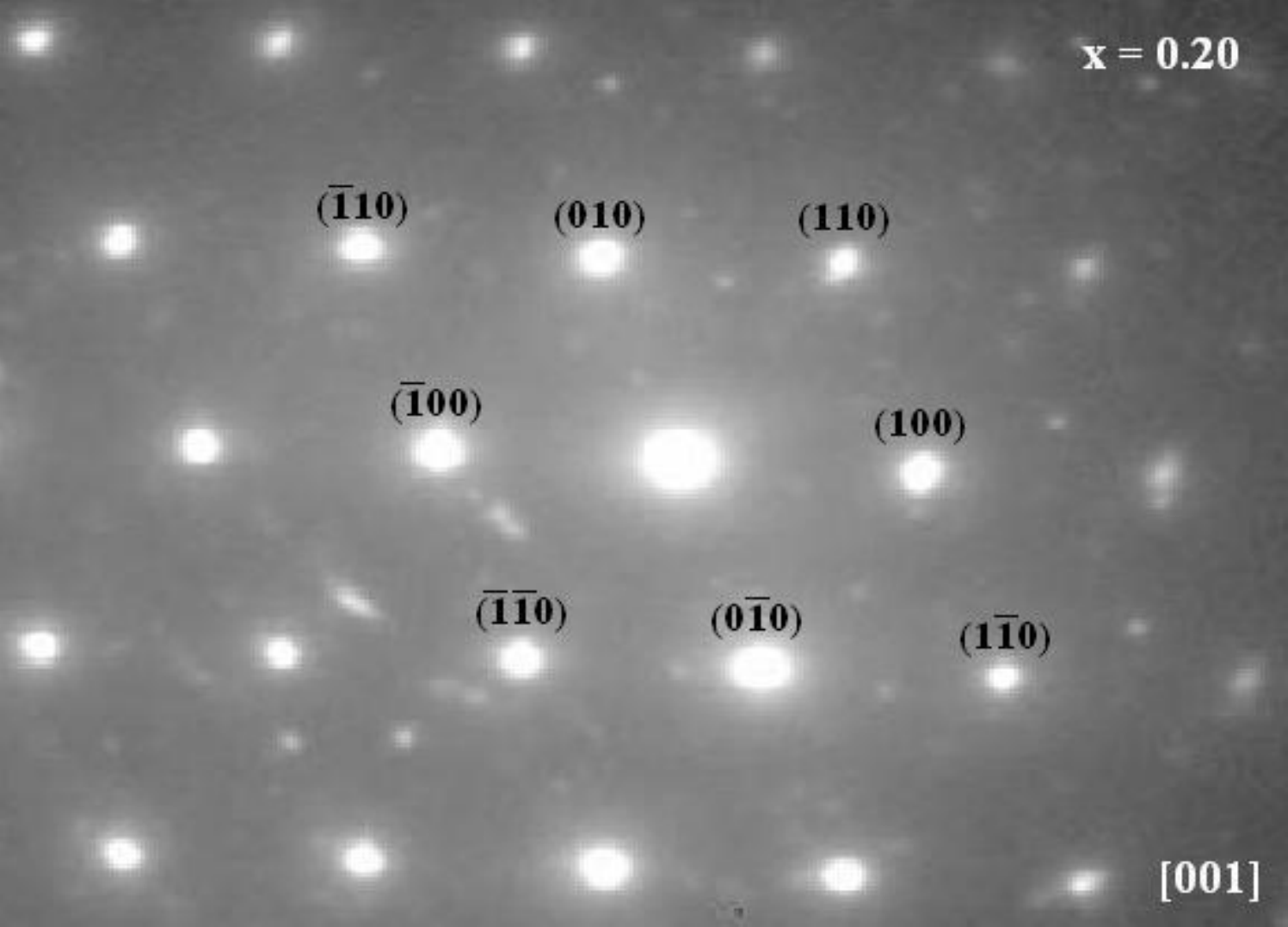}
  \end{center}

  \caption{\small Selected area electron diffraction (SAED) pattern
corresponding to figure (15) along [001] direction.}
  \label{fig-label}
\end{figure}
%%%%%%%%%%%%%%%%%%% Figure 17 %%%%%%%%%%%%%%%%%%%%%%%%%%%%%%%%%%%
\begin{figure}[htbp]
  \begin{center}
    \includegraphics[width=8cm,height=5.5cm,angle=0]{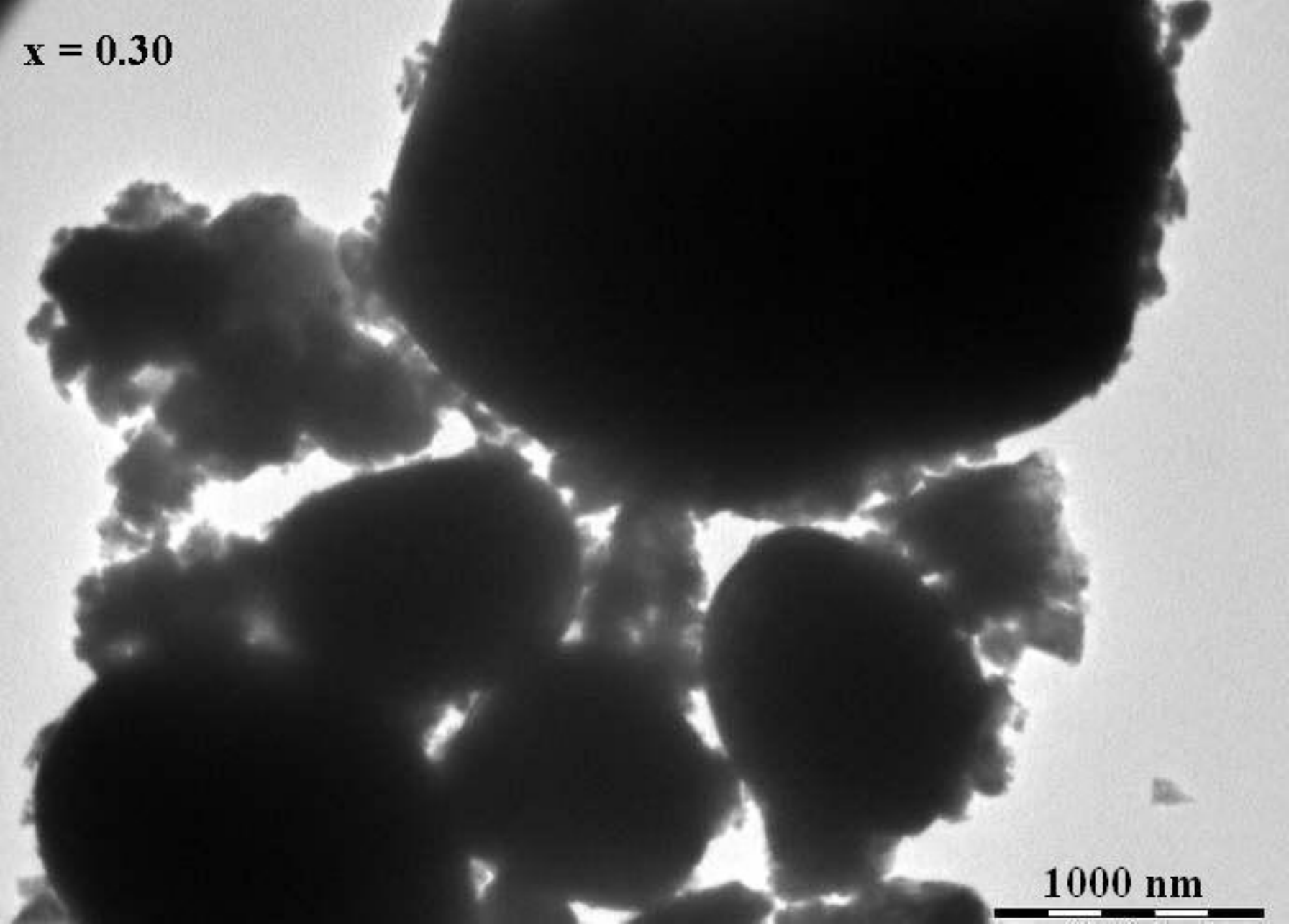}
  \end{center}

  \caption{\small Transmission electron micrograph of the bulk
$Sn_{1-x}Sb_{x}O_{2-\delta}$ system with x = 0.30 showing several
microcrystals.}
  \label{fig-label}
\end{figure}
%%%%%%%%%%%%%%%%%%% Figure 18 %%%%%%%%%%%%%%%%%%%%%%%%%%%%%%%%%%%
\begin{figure}[htbp]
  \begin{center}
    \includegraphics[width=8cm,height=5.5cm,angle=0]{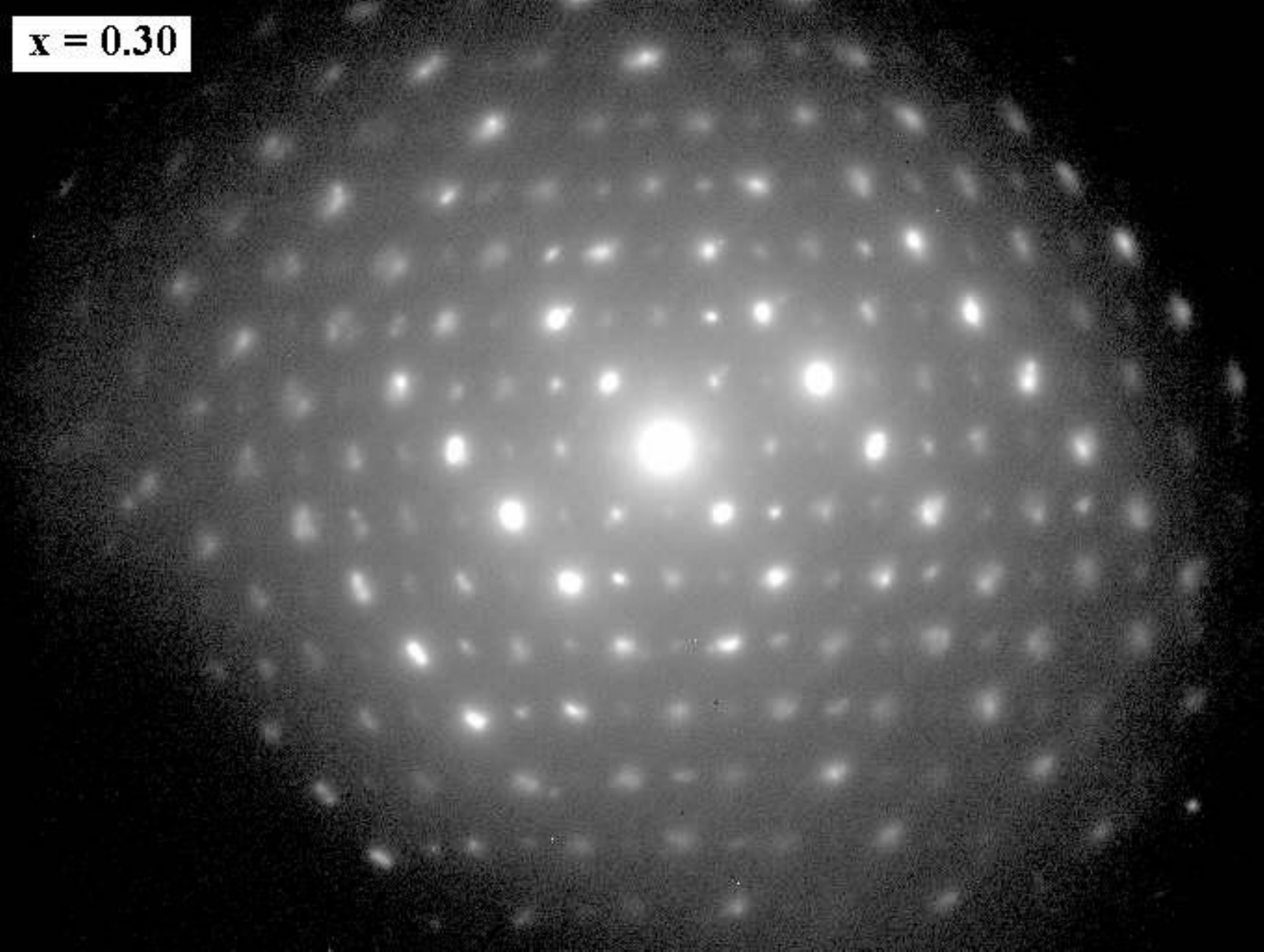}
  \end{center}

  \caption{\small Transmission electron micrograph of the bulk
$Sn_{1-x}Sb_{x}O_{2-\delta}$ system with x = 0.30 showing several
microcrystals.}
  \label{fig-label}
\end{figure}
%%%%%%%%%%%%%%%%%%%%%%%%%%%%%%%%%%%%%%%%%%%%%%%%%%%%%%%%%%%%%%%%%%%%
\section{Conclusions}
%%%%%%%%%%%%%%%%%%%%%%%%%%%%%%%%%%%%%%%%%%%%%%%%%%%%%%%%%%%%%%%%%%%%%%%%%%%%%
We have successfully prepared samples of type
$(Sn_{1-x}Sb_{x}O_{2-\delta})$ with x = 0.00, 0.10, 0.20 and 0.30 by
standard ceramic method. The analysis of x-ray diffraction patterns
revealed that the as synthesized doped and undoped tin oxides are
pure crystalline tetragonal rutile phase of tin oxide (JCPDS card
no. 041-1445) which belongs to the space group $P4_{2}/mnm$ (number
136). A small increase in the lattice parameters of the tetragonal
unit cell has been observed with increasing Sb content. This
possibly occurs due to the difference in ionic radii of $Sn^{4+}
(0.72{\AA})$ and $Sb^{3+} (0.90{\AA})$ ions. The average grain size
for the undoped tin oxide sample was found to be $\sim 67$ nm as
calculated by XRD using Debye Scherrer formula.\\\\Surface
morphology examination with SEM in scanning mode revealed the fact
that the grains are closely packed and pores between the grains are
in few numbers. These pores/voids between the grains increases with
antimony concentration up to 0.30. These electron micrographs also
reveal that the grain size in the antimony doped sample is larger
than that of undoped one.\\\\TEM image of undoped sample
indicates that the $SnO_{2}$ grains have diameters ranging from 25
to 120 nm and most grains are in cubic or spherical shaped. As
antimony content increases, the nanocubes/spheres are converted into
microcubes/spheres.\\\\Both reflectance and transmittance
of $Sn_{1-x}Sb_{x}O_{2-\delta}$ samples increases whereas absorbance
of these samples decreases with the increased concentration of
antimony (Sb) for the wavelength range 360 - 800 nm. The
energy bandgap of Sb doped - $SnO_{2}$ samples were obtained from
optical absorption spectra by UV-Vis absorption spectroscopy. Upon
increasing the Sb concentration the bandgap of the samples was found
to increase from 3.367 eV to 3.558 eV.\\\\
\textbf{Acknowledgement}\\\\
\small{The authors gratefully acknowledge Dr. Alok Kumar, GSI Lucknow for
providing facility of XRD and Dr. Jitendra Kumar of IIT Kanpur for
his help in recording the optical spectra. Authors also thank Dr. N.
P. Lalla and Dr. D. M. phase of UGC-DAE Consortium for Scientific
Research Center, Indore for TEM and SEM observations. Finally, we
would like to thank Prof. M. K. Mishra, Vice-chancellor and Prof. U.
D. Mishra, Head, Department of Physics, University of Lucknow for
their encouragement and support for providing the facilities for
material synthesis.}

%%%%%%%%%%%%%%%%%%% Figure 19 %%%%%%%%%%%%%%%%%%%%%%%%%%%%%%%%%%%
\begin{figure}[htbp]
  \begin{center}
    \includegraphics[width=8.5cm,height=6cm,angle=0]{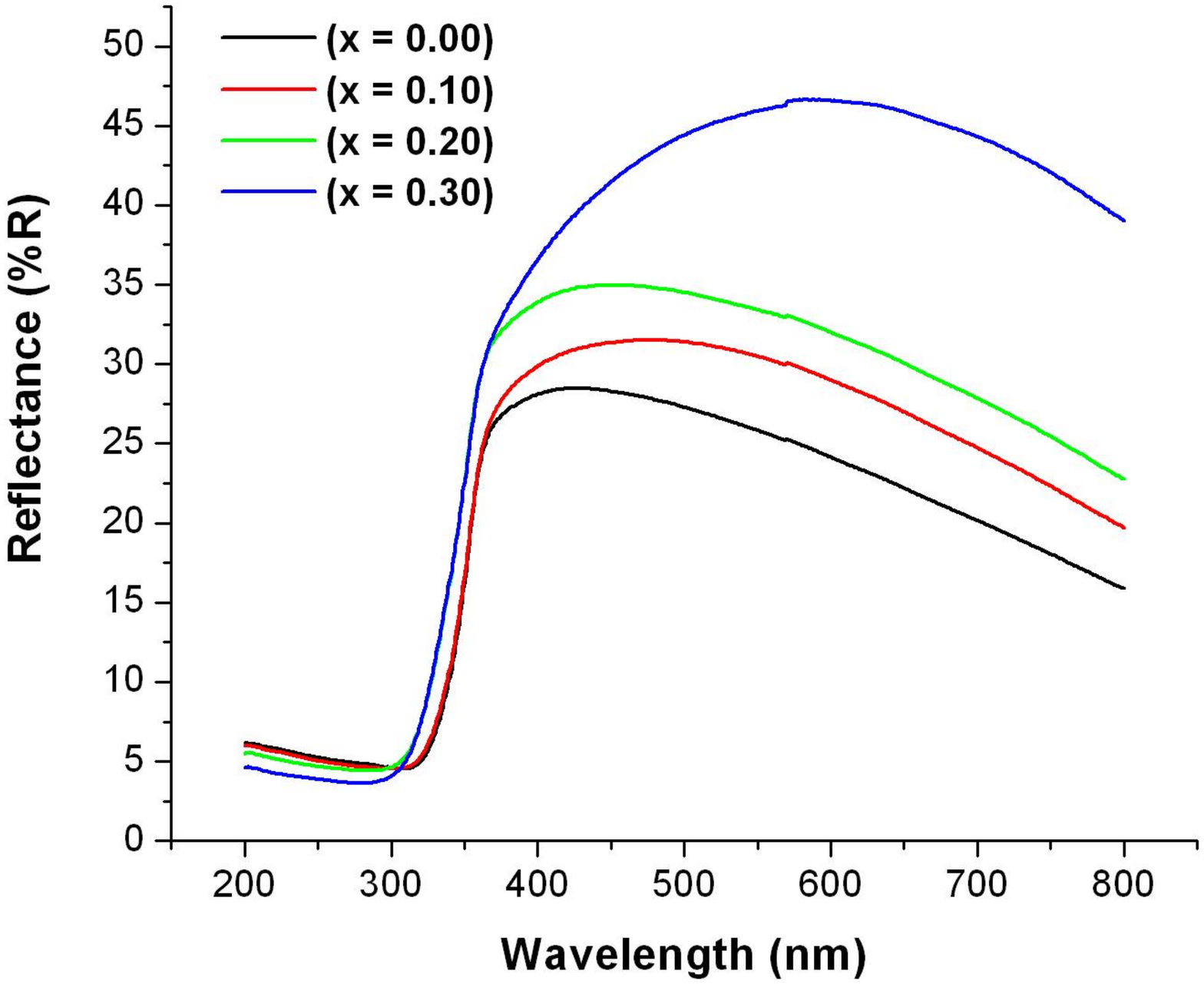}
  \end{center}

  \caption{\small UV-Vis reflectance spectra of
$Sn_{1-x}Sb_{x}O_{2-\delta}$ system with x = 0.00, 0.10, 0.20 and
0.30.}
  \label{fig-label}
\end{figure}
%%%%%%%%%%%%%%%%%%%%%%%%%%%%%%%%%%%%%%%%%%%%%%%%%%%%%%%
%%%%%%%%%%%%%%%%%%% Figure 20 %%%%%%%%%%%%%%%%%%%%%%%%%%%%%%%%%%%
\begin{figure}[htbp]
  \begin{center}
    \includegraphics[width=8.5cm,height=6cm,angle=0]{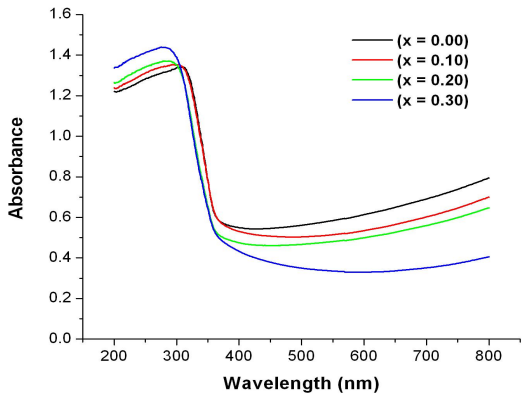}
  \end{center}

  \caption{\small UV-Vis absorbance Spectra of
$Sn_{1-x}Sb_{x}O_{2-\delta}$ system with x = 0.00, 0.10, 0.20 and
0.30.}
  \label{fig-label}
\end{figure}
%%%%%%%%%%%%%%%%%%%%%%%%%%%%%%%%%%%%%%%%%%%%%%%%%%%%%%%%%%%%%%%%%%%%%%
%%%%%%%%%%%%%%%%%%% Figure 21 %%%%%%%%%%%%%%%%%%%%%%%%%%%%%%%%%%%
\begin{figure}[htbp]
  \begin{center}
    \includegraphics[width=8.5cm,height=6cm,angle=0]{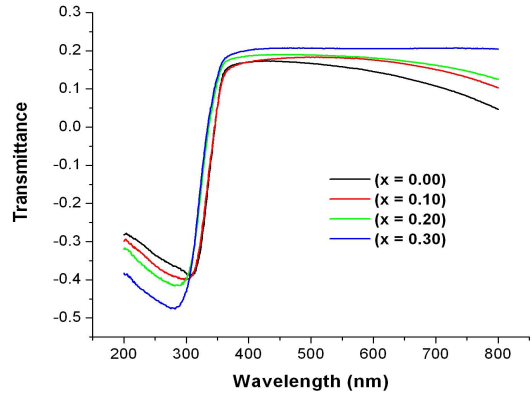}
  \end{center}

  \caption{\small Optical Transmittance spectra (calculated from
equation 4) of $Sn_{1-x}Sb_{x}O_{2-\delta}$ system with x = 0.00,
0.10, 0.20 and 0.30.}
  \label{fig-label}
\end{figure}
%%%%%%%%%%%%%%%%%%%%%%%%%%%%%%%%%%%%%%%%%%%%%%%%%%%%%%%
%%%%%%%%%%%%%%%%%%% Figure 22 %%%%%%%%%%%%%%%%%%%%%%%%%%%%%%%%%%%
\begin{figure}[htbp]
  \begin{center}
    \includegraphics[width=8.5cm,height=6cm,angle=0]{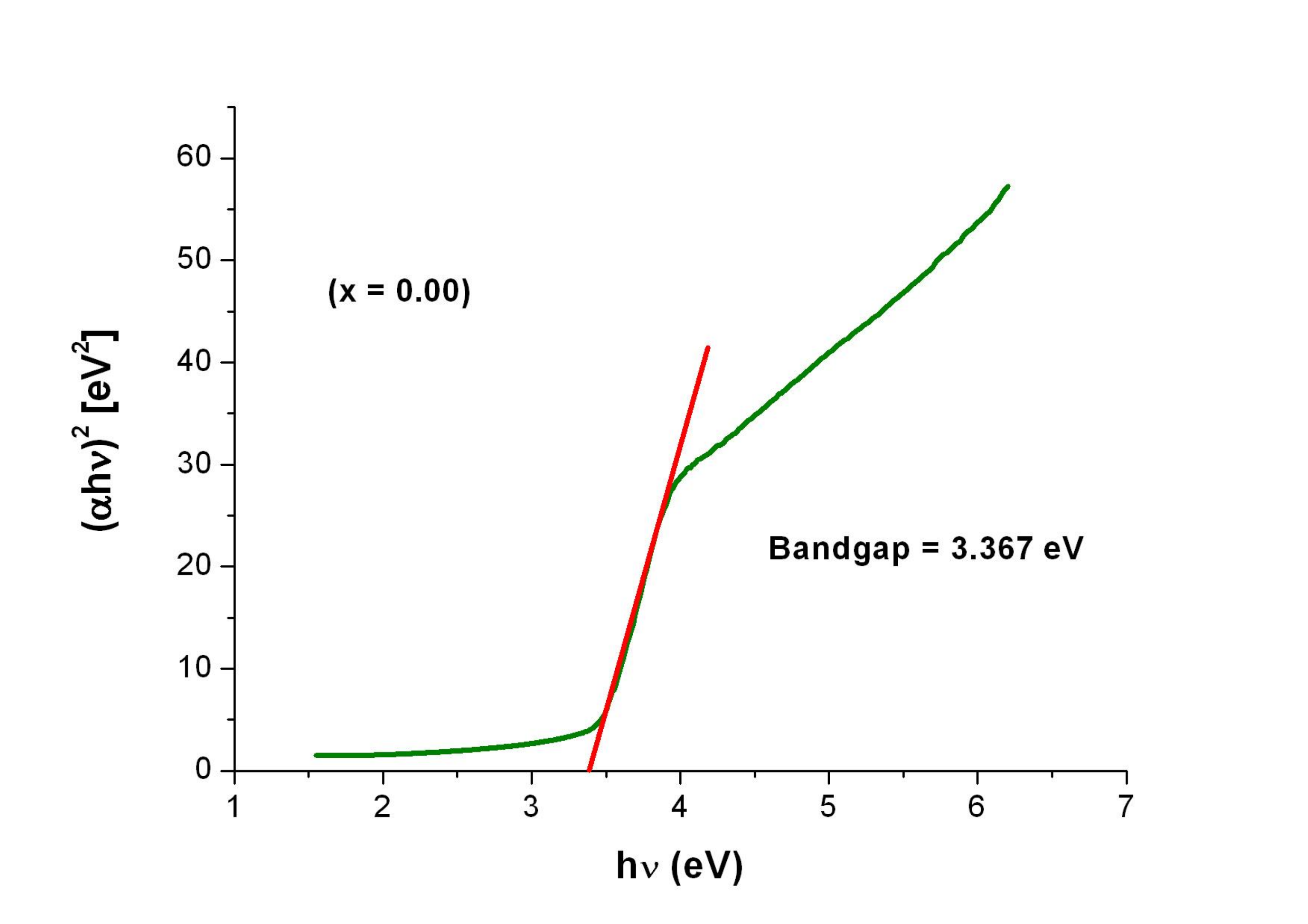}
  \end{center}

  \caption{\small $(\alpha h\nu)^2 [eV^{2}]$ versus photon energy
$(h\nu) [eV]$ curve for the bulk $Sn_{1-x}Sb_{x}O_{2-\delta}$ system
with x = 0.00. The direct energy bandgap $E_{g}$ is obtained from
the extrapolation to $\alpha = 0$.}
  \label{fig-label}
\end{figure}
%%%%%%%%%%%%%%%%%%% Figure 23 %%%%%%%%%%%%%%%%%%%%%%%%%%%%%%%%%%%
\begin{figure}[htbp]
  \begin{center}
    \includegraphics[width=8.5cm,height=6cm,angle=0]{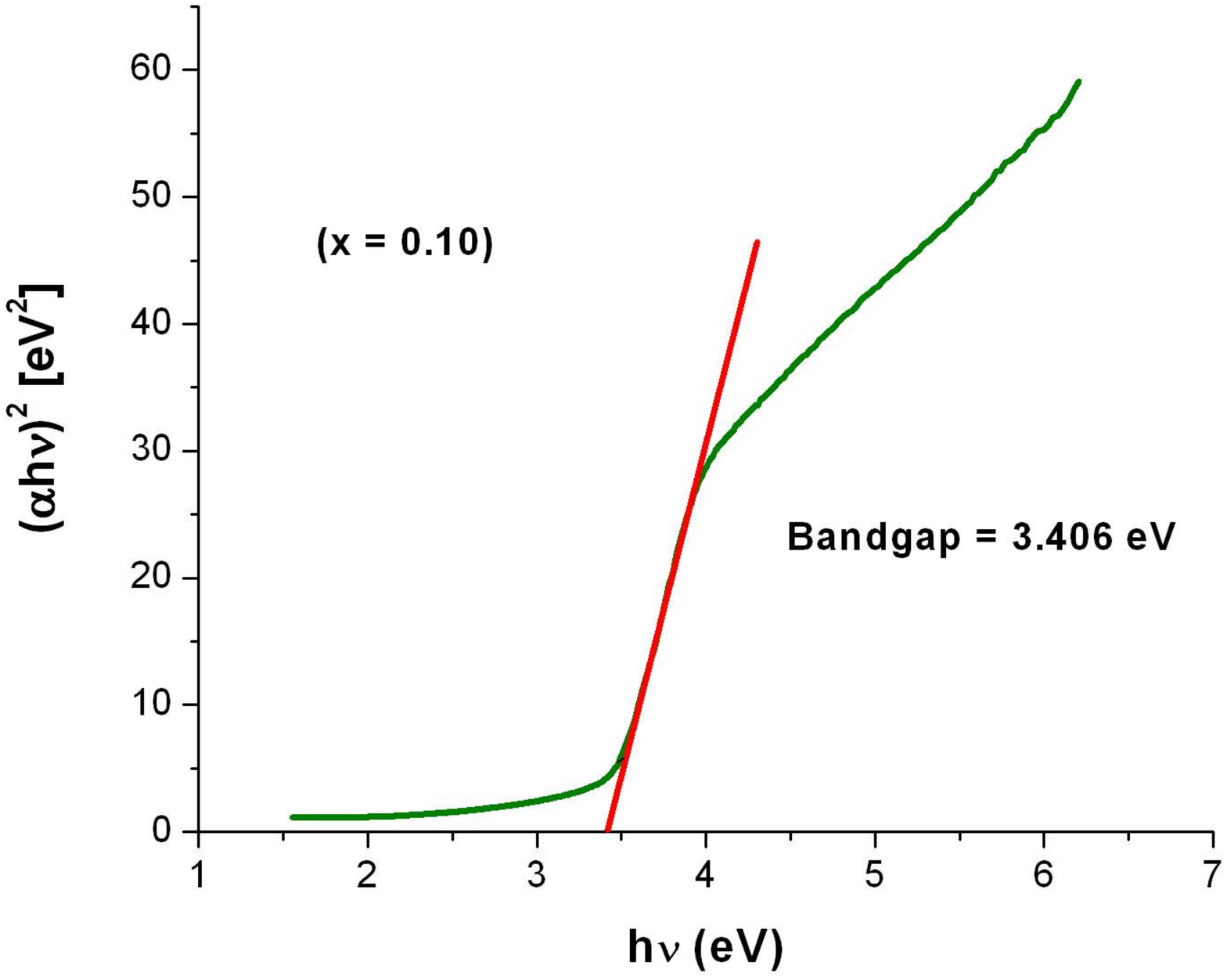}
  \end{center}

  \caption{\small $(\alpha h\nu)^2$ $[eV^{2}]$ versus photon energy
$(h\nu)$ [eV] curve for the bulk $Sn_{1-x}Sb_{x}O_{2-\delta}$ system
with x = 0.10. The direct energy bandgap $E_{g}$ is obtained from
the extrapolation to $\alpha = 0$.}
  \label{fig-label}
\end{figure}
%%%%%%%%%%%%%%%%%%% Figure 24 %%%%%%%%%%%%%%%%%%%%%%%%%%%%%%%%%%%
\begin{figure}[htbp]
  \begin{center}
    \includegraphics[width=8.5cm,height=6cm,angle=0]{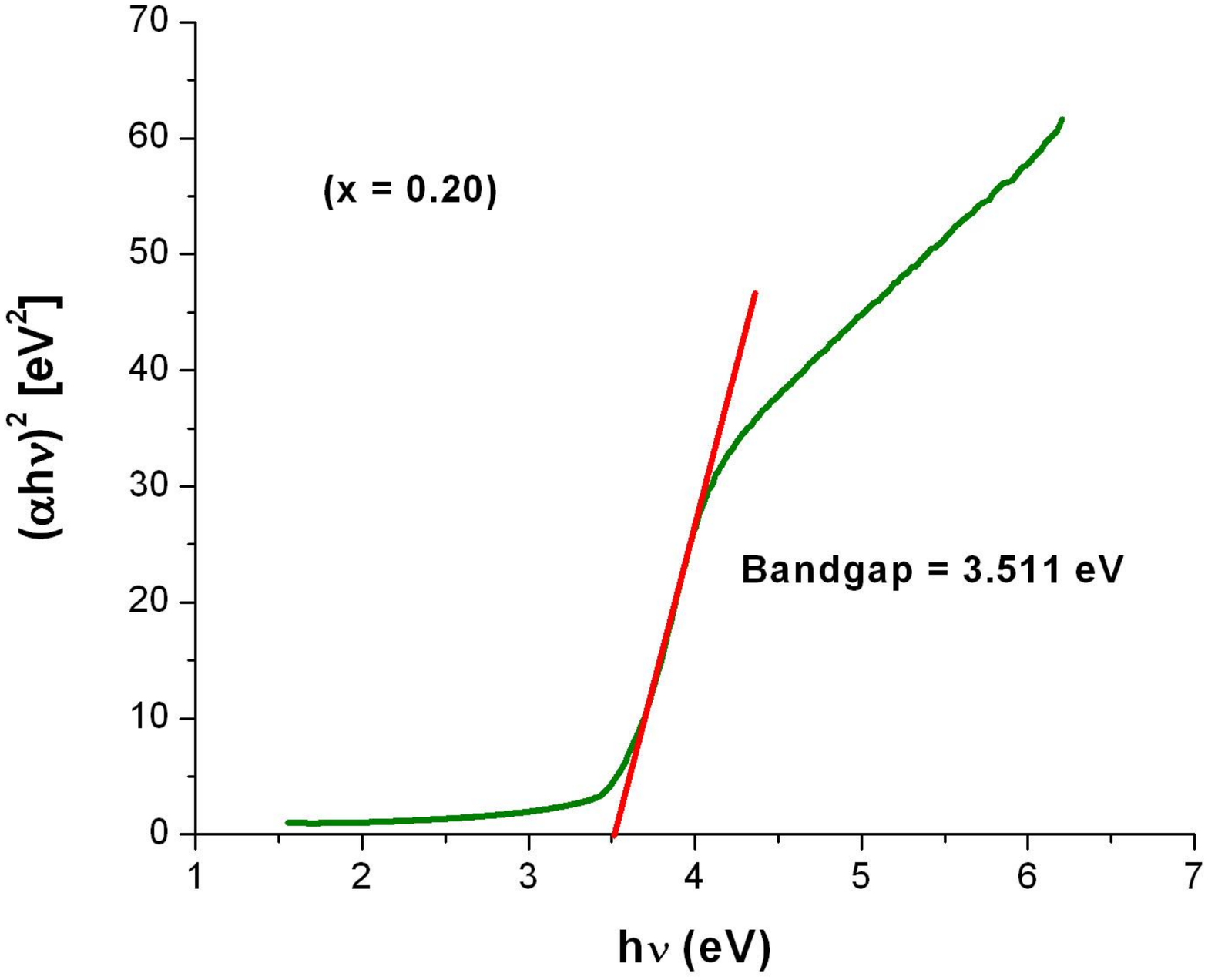}
  \end{center}

  \caption{\small $(\alpha h\nu)^2$ $[eV^{2}]$ versus photon energy
$(h\nu)$ [eV] curve for the bulk $Sn_{1-x}Sb_{x}O_{2-\delta}$ system
with x = 0.20. The direct energy bandgap $E_{g}$ is obtained from
the extrapolation to $\alpha = 0$.}
  \label{fig-label}
\end{figure}
%%%%%%%%%%%%%%%%%%%%%%%%%%%%%%%%%%%%%%%%%%%%%%%%%%%%%%%%%%%%%%%%%%%%%%%%%%%
\begin{figure}[htbp]
  \begin{center}
    \includegraphics[width=8.5cm,height=6cm,angle=0]{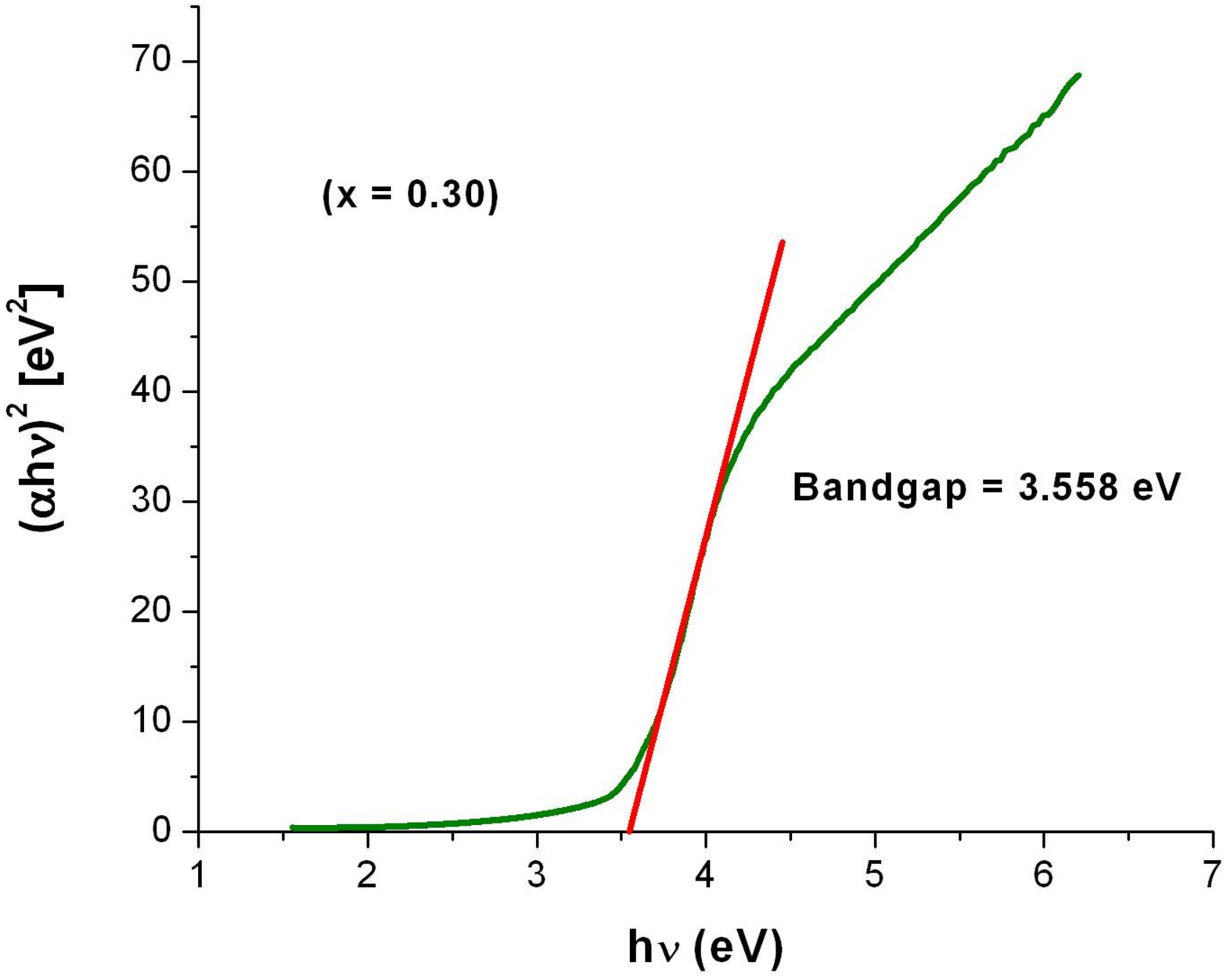}
  \end{center}

  \caption{\small $(\alpha h\nu)^2$ $[eV^{2}]$ versus photon energy
$(h\nu)$ [eV] curve for the bulk $Sn_{1-x}Sb_{x}O_{2-\delta}$ system
with x = 0.30. The direct energy bandgap $E_{g}$ is obtained from
the extrapolation to $\alpha = 0$.}
  \label{fig-label}
\end{figure}
%%%%%%%%%%%%%%%%%%%%%%%%%%%%%%%%%%%%%%%%%%%%%%%%%%%%%%%%%%%%%%%%%%
%%%%%%%%%%%%%%%%%%% Figure 26 %%%%%%%%%%%%%%%%%%%%%%%%%%%%%%%%%%%
\begin{figure}[htbp]
  \begin{center}
    \includegraphics[width=8cm,height=7cm,angle=0]{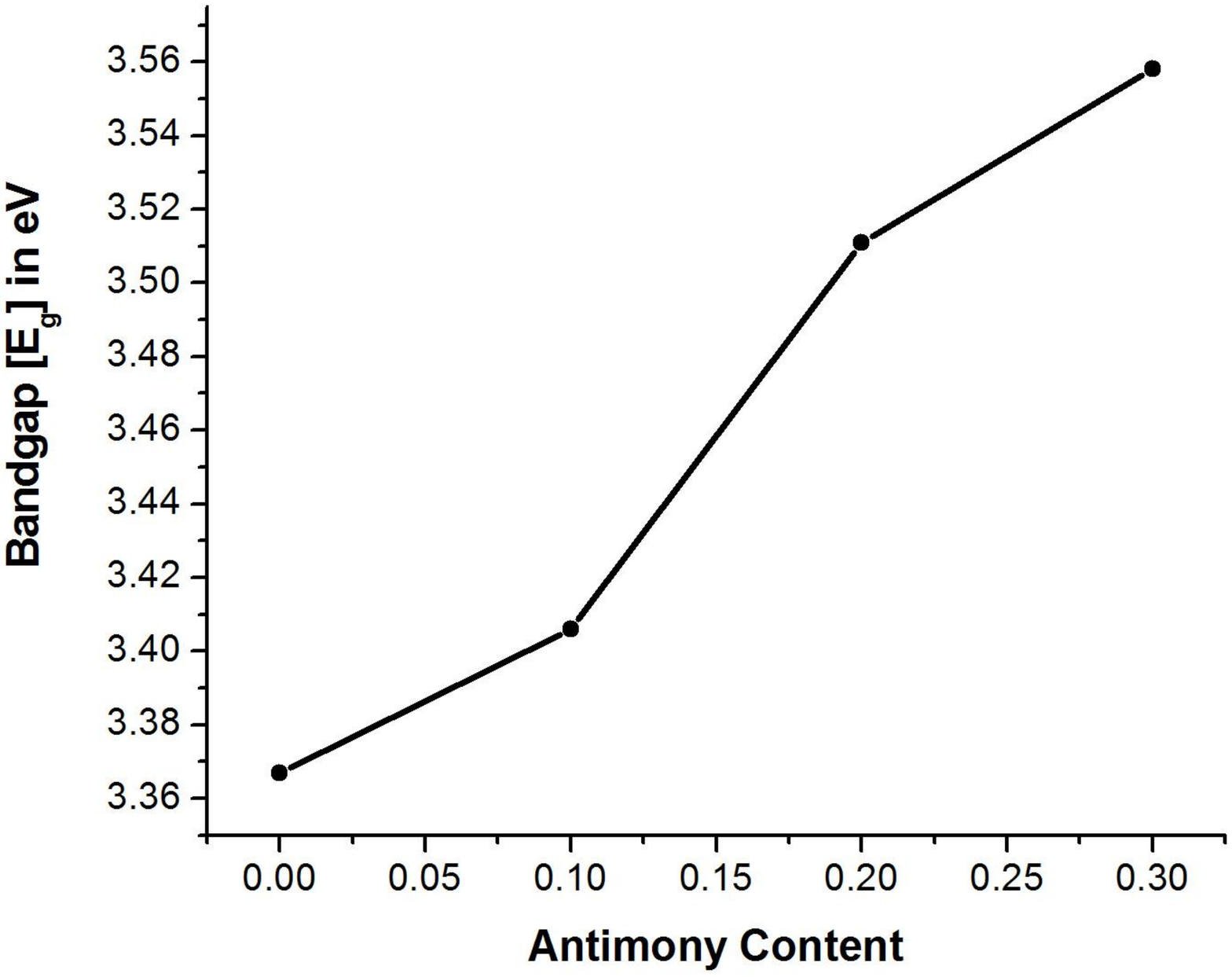}
  \end{center}

  \caption{\small Variation of bandgap $(E_{g})$ as a function of
increasing antimony (Sb) concentration. The bandgap of the samples
increased from 3.367 to 3.558 eV due to Sb doping.}
  \label{fig-label}
\end{figure}

\begin{thebibliography}{99}
\small{
\bibitem{ref1}
R. W. G. Wyckoff, Inorganic Compounds $RX_{n}$, $R_{n}MX_{2}$,
$R_{n}MX_{3}$, 2nd ed. (Interscience Publishers, New York, 1964).
\bibitem{ref2}
A. F. Wells, Structural Inorganic Chemistry, 6th ed. (Oxford
University Press, Oxford, UK, 1987).
\bibitem{ref3}
W. A. Harrison, Electronic Structure and the Properties of Solids:
the Physics of the Chemical Bond (Dover Publications, New York,
1989).
\bibitem{ref4}
H. H. Kung, Transition Metal Oxides: Surface Chemistry and Catalysis
(Elsevier, Amsterdam, Netherlands, 1989).
\bibitem{ref5}
V. E. Henrich and P. A. Cox, The Surface Science of Metal Oxides
(Cambridge University Press, Cambridge, UK, 1994).
\bibitem{ref6}
C. Noguera, Physics and Chemistry at Oxide Surfaces (Cambridge
University Press, 1996).
\bibitem{ref7}
J. P. Colinge, K. Hashimoto, T. Kamins, S. Y. Chiang, E. D. Liu, S.
S. Peng, and P. Rissman, Ieee Electron Device Lett. {\bf 7}, 279
(1986).
\bibitem{ref8}
V. Srinivasan and J. W. Weidner, J. Electrochem. Soc. {\bf 144},
L210 (1997).
\bibitem{ref9}
J. Watson, Sens Actuators {\bf 5}, 29 (1984).
\bibitem{ref10}
B. Noheda, Curr. Opin. Solid-State Mater. Sci. {\bf 6}, 27 (2002).
\bibitem{ref11}
K. Eguchi, T. Setoguchi, T. Inoue, and H. Arai, Solid State Ionics
{\bf 52}, 165 (1992).
\bibitem{ref12}
C. C. Hsu and N. L. Wu, J. Photochem. Photobiol. A {\bf 172}, 269
(2005).
\bibitem{ref13}
S. Lindroos and M. Leskela, Int. J. Inorg. Mater. {\bf 2}, 197
(2000).
\bibitem{ref14}
D. Golberg, Y. Bando, K. Fushimi, M. Mitome, L. Bourgeois, and C. C.
Tang, J. Phys. Chem. B {\bf 107}, 8726 (2003).
\bibitem{ref15}
K. Nakagawa, T. Nakata, and R. Konaka, J. Org. Chem. {\bf 27}, 1597
(1962).
\bibitem{ref16}
K. Nakagawa and T. Tsuji, Chem. Pharm. Bull. {\bf 11}, 296 (1963).
\bibitem{ref17}
K. Nakagawa, J. Sugita, and K. Igano, Chem. Pharm. Bull. {\bf 12},
403 (1964).
\bibitem{ref18}
K. Nakagawa, J. Sugita, and H. Onoue, Chem. Pharm. Bull. {\bf 12},
1135 (1964).
\bibitem{ref19}
C. Pierlot, V. Nardello, J. Schrive, C. Mabille, J. Barbillat, B.
Sombret, and J. M. Aubry, J. Org. Chem. {\bf 67}, 2418 (2002).
\bibitem{ref20}
C. G. Levi, J. Y. Yang, B. J. Dalgleish, F. W. Zok, and A. G. Evans, J.
Am. Ceram. Soc. {\bf 81}, 2077 (1998).
\bibitem{ref21}
A. Nazeri, P. P. TrzaskomaPaulette, and D. Bauer, J. Sol-Gel Sci.
Technol. {\bf 10}, 317 (1997).
\bibitem{ref22}
Z. R. Tian, W. Tong, J. Y. Wang, N. G. Duan, V. V. Krishnan, and S. L.
Suib, Science {\bf 276}, 926 (1997).
\bibitem{ref23}
I. E. Wachs, Catal. Today {\bf 27}, 437 (1996).
\bibitem{ref24}
G. C. Bond and S. F. Tahir, Appl. Catal. {\bf 71}, 1 (1991).
\bibitem{ref25}
Z. L. Wang, Annu. Rev. Phys. Chem. {\bf 55}, 159 (2004).
\bibitem{ref26}
M. Fernandez-Garcia, A. Martinez-Arias, J. C. Hanson, and J. A.
Rodriguez, Chem. Rev. {\bf 104}, 4063 (2004).
\bibitem{ref27}
M. E. Franke, T. J. Koplin, and U. Simon, Small {\bf 2}, 36 (2006).
\bibitem{ref28}
N. Barsan, D. Koziej, and U. Weimar, Sens. Actuators B {\bf 121}, 18
(2007).
\bibitem{ref29}
P. Barbarat, S. F. Matar, and G. LeBlevennec, J. Mater. Chem. {\bf
7}, 2547 (1997).
\bibitem{ref30}
M. Calatayud, J. Andres, and A. Beltran, Surf. Sci. {\bf 430}, 213
(1999).
\bibitem{ref31}
L. A.  Errico, Physica B {\bf 389}, 140 (2007).
\bibitem{ref32}
M. A. Maki-Jaskari and T. T. Rantala, Phys. ReV. B {\bf 64}, 075407
(2001).
\bibitem{ref33}
J. Robertson, K. Xiong, and S. J. Clark, Thin Solid Films {\bf 496},
1 (2006).
\bibitem{ref34}
D. Maestre, J. Ramirez-Castellanos, P. Hidalgo, A. Cremades, J. M.
Gonzalez-Calbet, and J. Piqueras, Eur. J. Inorg. Chem., 1544 (2007).
\bibitem{ref35}
F. Ming, T. Xiaoli, C. Xueli, Z. Lide, L. Peisheng, and J. Zhi, J.
Phys. D: Appl. Phys. {\bf 40}, 7648 (2007).
\bibitem{ref36}
C. A. Vincent, J. Electrochem. Soc. {\bf 119}, 515 (1972).
\bibitem{ref37}
B. Alterkop, N. Parkansky, S. Goldsmith, and R. L. Boxman, J. Phys. D
{\bf 36}, 552 (2003).
\bibitem{ref38}
S. Samson and C. G. Fonstad, J. Appl. Phys. {\bf 44}, 4618 (1973).
\bibitem{ref39}
J. Maier and W. J. Gopel, Solid State Chem. {\bf 72}, 293 (1988).
\bibitem{ref40}
P. Dinola, F. Morazzoni, R. Scotti, and D. Narducci, J. Chem. Soc.,
Faraday Trans. {\bf 89}, 3711 (1993).
\bibitem{ref41}
F. Morazzoni, C. Canevali, N. Chiodini, C. Mari, R. Ruffo, R.
Scotti, L. Armelao, E. Tondello, L. Depero, and E. Bontempi, Mater.
Sci. Eng., C {\bf 15}, 167 (2001).
\bibitem{ref42}
S. H. Sun, G. W. Meng, G. X. Zhang, T. Gao, B. Y. Geng, L. D. Zhang, and
J. Zuo, Chem. Phys. Lett. {\bf 376}, 103 (2003).
\bibitem{ref43}
R. Ningthoujam, D. Lahiri, V. Sudarsan, H. K. Poswal, S. K.
Kulshreshtha, S. M. Sharma, B. Bhushan, and M. D. Sastry, Mater. Res.
Bull. {\bf 42}, 1293 (2007).
\bibitem{ref44}
J. D. Prades, J. Arbiol, A. Cirera, J. R. Morante, M. Avella, L.
Zanotti, E. Comini, G. Faglia, and G. Sberveglieri, Sensor Actuat. B
{\bf 126}, 6 (2007).
\bibitem{ref45}
X. C. Yang, Mater. Sci. Eng. B {\bf 93}, 249 (2002).
\bibitem{ref46}
J. Tauc, R. Grogorovici and A. Vancu, Phys. Stat. Solidi. {\bf 15},
627 (1966).
\bibitem{ref47}
T. Arai, J. Phys. Soc. Japan {\bf 15}, 916 (1960).
\bibitem{ref48}
D. Frohlich and R. Kenklies, Phys. Rev. Lett. {\bf 41}, 1750 (1978).
\bibitem{ref49}
F. Gu, S. F. Wang, M. K. Lu, X. F. Cheng, S. W. Liu, G. J. Zhou, D. Xu,
and D. R. Yuan, J. Cryst. Growth {\bf 262}, 182 (2004).
\bibitem{ref50}
A. E. De Souza, S. H. Monteiro, C. V. Santilli, and S. H. Pulcinelli, J.
Mater. Sci.: Mater. Electron. {\bf 8}, 265 (1997).
\bibitem{ref51}
E. R. Leite, M. Ines, B. Bernardi, E. Longo, J. A. Varela, and C. A.
Paskocimas, Thin Solid Films {\bf 449}, 67 (2004).
\bibitem{ref52}
S. Tsunekawa, T. Fukuda, and A. Kasuya, J. Appl. Phys. {\bf 87},
1318 (2000).
\bibitem{ref53}
E. Burstein, Phys. Rev. {\bf 93}, 632 (1954).
\bibitem{ref54}
T. S. M\"{o}ss, Proc. Phys. Soc. Lond. Ser. B {\bf 67}, 775 (1954).}
\end{thebibliography}
\end{document}